\newif\ifTCOMorTWCDraft
\TCOMorTWCDraftfalse

\ifTCOMorTWCDraft
    \documentclass[12pt, draftclsnofoot, onecolumn]{IEEEtran}
\else
    \documentclass[10pt]{IEEEtran}
\fi

\usepackage[utf8]{inputenc}
\usepackage{amsmath}
\usepackage{amssymb}
\usepackage[makeroom]{cancel}
\usepackage{graphicx, booktabs, array}
\usepackage{amsthm}
\usepackage{xcolor}
\usepackage{xargs}   % Use more than one optional parameter in a new commands
\usepackage{xcolor}  % Coloured text etc.
\usepackage{algorithm}
\usepackage{algpseudocode}

\usepackage{tikz}
\usepackage{pgfplots} 
\usepackage{pgfgantt}
\usepackage{pdflscape}
\pgfplotsset{compat=1.16, plot coordinates/math parser=true}
\pgfplotsset{every tick label/.append style={font=\footnotesize}}
\pgfplotsset{every axis/.append style={label style={font=\footnotesize}, width=7cm, height=5.5cm}}

\usepackage{listings}
\usepackage{color} %red, green, blue, yellow, cyan, magenta, black, white
\definecolor{mygreen}{RGB}{28,172,0} % color values Red, Green, Blue
\definecolor{mylilas}{RGB}{170,55,241}
\usepackage{authblk}
\usepackage{hyperref}

\newtheorem{theorem}{Theorem}
\newtheorem{corollary}[theorem]{Corollary}

\newtheorem{lemma}[theorem]{Lemma}
\newtheorem{definition}{Definition}
\newtheorem{proposition}[theorem]{Proposition}

\newcommand{\norm}[1]{\left\lVert#1\right\rVert}

\title{Order-optimal Joint Transmission and Identification in Massive Multi-User MIMO via Group Testing}
\author{George Vershinin, Asaf Cohen, Omer Gurewitz}
\affil{The School of Electrical and Computer Engineering\\
Ben-Gurion University of the Negev}

\date{August 2022}
\begin{document}
\maketitle
\begin{abstract}
    The number of wireless devices which are connected to a single Wireless Local Area Network continues to grow each year. As a result, the orchestration of so many devices becomes a daunting, resource--consuming task, especially when the resources available at the single access point are limited, and it is hard to anticipate which devices will request access at any given time. On the other hand, the number of antennas on both the devices and the access point grows as well, facilitating advanced joint scheduling and coding techniques.
    
    In this paper, we leverage the large number of antennas and suggest a massive multiple-user multiple-input-multiple-output (MU-MIMO) scheme using sparse coding based on Group Testing (GT) principles. The scheme allows for a small subset of devices to transmit simultaneously, without a preceding scheduling phase or coordination, thus reducing overhead and complexity.
    Specifically, we show that out of a population of \(N\) devices, it is possible to jointly identify and decode \(K\) devices, unknown in advance, simultaneously and without any scheduling. The scheme utilizes minimal knowledge of channel state, uses an efficient (in both run-time and space) decoding algorithm, and requires \(O(K\log N\mathcal{M})\) antennas, where \(\mathcal{M}\) is the number of messages per device. In fact, we prove that this scheme is order--optimal in the number of users and messages.
    This is done by deriving sufficient conditions for a vanishing error probability (a direct result), bounding the minimal number of antennas necessary for any such scheme (a converse result), and showing that these results are asymptotically tight.
\end{abstract}

\section{Introduction}
Multiple-input-multiple-output (MIMO) systems have become ubiquitous due to their increased reception and transmission quality, in both single and multi-user (MU) communications.
In MU communication, most MU-MIMO works focus on user selection (e.g., \cite{Xia2021approx_KKT_iterative, Zhang2021clusteredJointPowerOpt}) as a possible solution to the Multiple Access Channel (MAC) problem.
Even the 802.11ax standard, the state-of-the-art WiFi technology, solves the MAC problem by scheduling users to dedicated frequency bands \cite[Chapters~3.3.4-3.3.6]{gulasekaran2022wifi6}, scheduling only a very small group of users simultaneously.
Traditional user selection carries a lot of overhead - complex optimization problems solved by a \emph{centralized unit}, information gathering (by message passing), and the scheduling announcement.
The announcements may use dedicated resources - reducing system resource efficiency.
Collecting and processing Channel State Information (CSI) to schedule users can be computationally hard when the number of users is large, so optimal user scheduling is infeasible.
Reducing this complexity encompasses many challenges;
Moving the scheduling task to the users by means of self-scheduling requires sophisticated mechanisms to identify them and their transmitted codewords.
Additionally, when no CSI is present, the receiver is greatly limited in its processing options.
E.g., using matched filters.

We address these challenges by combining two seemingly unrelated ideas into a novel, order-optimal self-scheduling, identification, and decoding scheme.
The first is Index Modulation, and the other is Group Testing (GT).
In Index Modulation, users selectively activate antennas at the receiver to send information \cite[Chapter~1.2]{IndexModBook} in addition to transmitting a symbol.
In such a scheme, it is easy to construct an algorithm that identifies the transmitting user by the activated antenna's index.
The symbol can be recovered in numerous ways.
For example, using a single threshold like in On-Off Keying \cite{enrgyDetectSISO}.

Then, we revisit the GT problem, capable of finding $K$ ill patients (or defective items) out of a large population of $N$ patients \cite{GTBook}, in order to devise a mechanism to identify the senders and decode their information.
Indeed, in GT, patients are tested together rather than individually, minimizing the number of tests required to identify the $K$ ill patients.
Specifically, in GT, the patients participating in each test can be determined \emph{a-priori} in the form of a test matrix.
After conducting all tests, the test conductor observes the result vector and uses decoding algorithms, such as Chan's Noisy Column Matching (Noisy CoMa) \cite{CoMa, CoMa2} to identify the defective items.

In the context of user scheduling and identification, the users and their messages are analogous to the population of $N$ items.
The $K$ items of interest are the self-scheduled users who actually send messages whose identity is unknown.
The test matrix is akin to a binary codebook.
The tests conducted are usually energy detection results on different system resources such as timeslots, frequency bands, or, as in the suggested scheme, antennas activated.

Modern literature suggests GT-originated codes to devise communication protocols capable of joint decoding many messages using a simple decoding algorithm.
For example, in \cite{robin2021sparse}, Robin and Erkip proposed an energy-efficient sensor discovery in power-constrained clustered networks.
Cohen \emph{et al.} proposed a GT-based communication protocol in \cite{EffiColl}, based on the binning ideas from \cite{SecureGroupTesting}.
Robin and Erkip have analyzed a protocol similar to \cite{EffiColl} in \cite{robin2021capacity}, where they assumed a Rayleigh Fading Channel.
The main idea is to reduce the continuous signal and noise models into discrete binary models, using energy detection, followed by Noisy CoMa for decoding.
The energy detection suggested straightforwardly compares the energy at the timeslot to some threshold.

These works assumed transmissions over timeslots and can be easily extended to frequency bands.
However, the extension to the spatial dimension, where antennas act as test tubes, is far from trivial;
Unlike different timeslots or frequency bands, transmissions from each antenna directly affect the signals at \emph{all} receiving antennas.
In other words, adjusting existing schemes for MU-MIMO requires careful design to prevent self-interference.

In the suggested scheme, users have a codebook generated using methods from GT.
The users leverage their massive number of antennas to null their transmitted signals' energy in the antennas corresponding to zeros in their allocated codeword, and allow energy at the antennas corresponding to ones.
The receiver uses energy detection (converting channel output to a binary vector) to estimate which antenna is targeted by at least one user.
The binary vector is treated as the result vector of GT and is the input to a decoding algorithm which returns the sent messages (consequently, the identities of the transmitting users as well).

Our scheme requires no scheduling overhead (headers, control messages, CSI collection at the receiver, etc.) and has extraordinarily low complexity;
Codeword length is linear in the number of antennas.
Their growth is logarithmic in the total number of users in the system and is linear with the number of self-scheduled users.
Each antenna is equipped with a simple energy detector which compares the input energy to some pre-defined energy threshold.
The decoding algorithm is efficient in both run-time and space, requiring only $O(N \mathcal{M} K\log N\mathcal{M})$ operations and \emph{no additional space} aside from trivially storing the codebook.
Our suggested scheme's space requirements significantly improve the $\Omega(N)$ space complexity required for optimal user scheduling.
We analyze the error probability of the scheme, find a lower bound and scaling laws of the number of antennas, and show that it is order-optimal in either the number of users or the number of messages per user.

The rest of the paper is organized as follows:
We introduce our notations and the model in Section \ref{sect:model}.
Section \ref{sect:sol_scheme} describes our GT-based scheme and discusses the results.
We thoroughly analyze our system in Section \ref{sect:scheme_analysis}.
In Section \ref{sect:converse}, we formulate a necessary lower bound (converse) on the number of antennas for a vanishing error probability.
In Section \ref{sect:sim_num_res}, we show numerical evaluations of our results and compare our scheme with existing technologies.
Section \ref{sect:conc} concludes the paper.

% \section{Related Work}
% \label{sect:relatedWork}
% \subsection{Group Testing}
% \label{subsect:GT}
% \input{TexFiles/02 - Related Work/RelatedWork_GT_Short}
% \subsection{MIMO}
% \label{subsect:MIMO}
% \input{TexFiles/02 - Related Work/RelatedWork_MIMO_Short}

\section{System Model}
\label{sect:model}
\subsection{Notation}
Matrices will appear in bold (e.g., $\mathbf{H}$) and vectors are underlined (e.g., $\underline{x}$).
We shall use subscript for user indices (e.g., $\mathbf{H}_i$), components of a vector or matrix are specified as a subscript after squared parentheses (e.g., $[\underline{y}]_m$ is $\underline{y}$'s $m^{\mathrm{th}}$ component, $[\mathbf{H}_k]_{i,j}$ is $\mathbf{H}_k$'s component in the $i^{\mathrm{th}}$ row and $j^{\mathrm{th}}$ column).
All logarithms in this article are in base two.
When they are not, we will specify the log base explicitly or write $\ln(\cdot)$ in the case of the natural logarithm.
We write $[n]$ as a shorthand notation for the set $\{1,2,\dots, n\}$.
We write $2^{\mathcal{S}}$ as the power set of a set $\mathcal{S}$ (e.g., $2^{[n]}$).
We use a single subscript after squared parentheses of a matrix to specify a column of a matrix.
E.g., $[\mathbf{H}_k]_j$ is $\mathbf{H}_k$'s $j^{\mathrm{th}}$ column.
We use $(\cdot)^T$ for the transpose operation and $(\cdot)^*$ for the Conjugate Transpose.
We write $\Re(X)$ and $\Im(X)$ to denote the real and imaginary parts of some complex variable $X$, respectively.
\subsection{Model}
We assume a time-slotted network of $N$ users where $K\ll N$ of them simultaneously transmit to a single receiver.
In each time slot, a different set of $K$ users may transmit, and their identity is unknown a-priori.
Every user wishes to send one out of $\mathcal{M}$ possible messages using a \emph{single} time slot and a \emph{single} frequency band, and there is no a-priori knowledge about the distribution of which a message is sent.
The $k^{\mathrm{th}}$ user's messages are $\mathcal{W}_k = \{w_{k,1}, w_{k,2}, \dots, w_{k,\mathcal{M}}\}$.
WLOG, the transmitting users are $[K]$, and each user wishes to transmit its first message, $w_{k,1}$.

Each transmitter has $M_t$ antennas, whereas the receiver has $M_r$ antennas.
Each transmitter has complete knowledge of its channel state at any given time but has no channel state of other transmitters (CSIT model, as named in \cite[Chapter~10]{goldsmith_2005}).
The receiver, on the other hand, has no CSI.
The channel matrix of the $k^{\mathrm{th}}$ user is $\mathbf{H}_k\in\mathbb{C}^{M_r\times M_t}$.
Each entry in $\mathbf{H}_k$ is a zero-mean Complex Gaussian Random Variable (CGRV).
We also assume a zero-mean White Complex Gaussian Additive Noise, $\underline{n}\in\mathbb{C}^{M_r\times1}$, where $[\underline{n}]_i\sim\mathcal{CN}(0, N_0)$ for all $i$.

Since $\mathbf{H}_k$ is known to the $k^{\mathrm{th}}$ transmitter, its encoder is a function that maps some $w_k\in\mathcal{W}_k$ to a complex vector $\underline{x}_k\in\mathbb{C}^{M_t\times 1}$.
The choice of $\underline{x}_k$ also depends on the CSI, for example, by beamforming.
Mathematically:
\begin{align}
    \mathcal{E}_k:\mathcal{W}_k\times\mathbb{C}^{M_r\times M_t}\to\mathbb{C}^{M_t\times 1}
\end{align}
The transmission cannot exceed some power level, $P$.
I.e., $\norm{\underline{x}_k}^2\leq P$.
For convenience, we shall assume each component of $\mathbf{H}_k$ has a unit variance\footnote{The assumption of unit variance at the channel matrix can be interpreted as the normalization of the power constraint with the fading power.
I.e., $P$ is not the transmitted power perse, but the average SNR per antenna.}, that is, $[\mathbf{H}_k]_{i,j}\sim\mathcal{CN}(0, 1)$ for all $i,j,k$.
Finally, the receiver obtains
\begin{gather}
    \label{eq:SysMod_y}
    \underline{y}
    =
    \sum_{k=1}^{K}\mathbf{H}_k\underline{x}_k + \underline{n}.
\end{gather}

The decoder uses $\underline{y}$ to obtain the messages sent and infer the identity of the $K$ users.
Hence, the decoder is a function
\begin{align}
    \mathcal{D}: \mathbb{C}^{M_r\times 1}\to \prod_{k=1}^N \mathcal{W}_k\times [N]^N.
\end{align}
The product taken in this case is the Cartesian Product.
Messages and identities of silent users are defined as $\emptyset$ and are omitted from the function output\footnote{
    This definition allows decoders to output \emph{any} number of messages up to $N$, assuming each user sends up to one message per time slot.
}.

We assume $M_r = M_t$, and a Massive MIMO settings, so $1\ll M_t, M_r$.
When a system has a minuscule number of antennas, one can use additional time slots/frequency bands to compensate for the difference.
E.g., if our solution requires $L = l\cdot M_r$ antennas, use $l$ time slots.
In each time slot, save the channel output to obtain $\{\underline{y}_i\}_{i=1}^{l}$.
Finally, $\underline{y}$ can be obtained by concatenation: $\underline{y} = (\underline{y}_1^T,\underline{y}_2^T,\dots,\underline{y}_l^T)^T$.

Throughout, we will give expressions valid for any $N$, $K$, and $\mathcal{M}$, yet we will focus on their relationship such that the error probability, defined below, will vanish.
% We are interested in the finite-block length regime and asymptotic behavior when either $N\to\infty$ or $C\to\infty$ for any $K$.
% $K$ is allowed to scale with $N$ (increasing the number of simultaneously transmitting users) or $C$.
We say that the system is \emph{message-user reliable} if
\begin{align}
    \mathbb{P}(
        \mathcal{D}(\underline{y})
        &\neq
        (\underbrace{w_{1,1}, w_{2,1},\dots, w_{K,1}}_{\text{The sent messages}},
        % \underbrace{\emptyset,\dots,\emptyset}_{\substack{\text{No}\\\text{messages}}},
        \underbrace{[K]}_{\text{Identities}})
    )
    \underset{N,\mathcal{M}\to\infty}{\xrightarrow{\hspace*{1cm}}}
    0.
\end{align}
That is, the decoder obtains \emph{exactly} $K$ messages, and the correct ones.
Additionally, it must correctly identify the corresponding users.
\subsection{Rates}
\label{subsect:Rates}
Each user has $\mathcal{M}$ codewords, hence sends $\log \mathcal{M}$ bits per transmission.
When identifying the $K$ users, the receiver decodes additional $\log\binom{N}{K}$ bits.
For an approximation for large $N$, replace $\log\binom{N}{K}$ with $K\log N$.
We thus obtain a sum-rate of $K\log \mathcal{M} + K\log N = K\log N\mathcal{M}$ bits per channel use.

We are interested in comparing this rate with the ergodic sum-rates.
The ergodic sum-rate has been established in \cite{MIMO_MAC} (albeit for models with CSI at the receiver).
They defined and used capacity notations to approximate the capacity of MU-MIMO systems to scaled (by $\min\{K M_t, M_r\}$) versions of single-user to single receiver ergodic capacity, where each party has exactly one antenna.
When $M_r < K M_t$, their approximation has errors bounded by a logarithmic (in $K M_t$) term.
Adapting their result, without the capacity notation, to our system model yields the ergodic sum-rate with CSIT \cite[Equation~(40)]{MIMO_MAC}:
\ifTCOMorTWCDraft
    \begin{align}
        \nonumber
        C_{Full\ CSI}
        &\approx
        M_r \cdot \mathbb{E}[
            \log (1+\rho |[\mathbf{H}_1]_{1,1}|^2)
        ]
        +
        O(
            \log(1+K M_t-M_r)
        )
        \\\label{eq:ergodic_Cap_K_CSIT}
        &\leq
        M_r \log(1+\rho)
        +O(
            \log(K M_r)
        )
    \end{align}
\else
    \begin{align}
        \nonumber
        C_{Full\ CSI}
        &\approx
        M_r \cdot \mathbb{E}[
            \log (1+\rho |[\mathbf{H}_1]_{1,1}|^2)
        ]
        \\\nonumber
        &\qquad
        +
        O(
            \log(1+K M_t-M_r)
        )
        \\\label{eq:ergodic_Cap_K_CSIT}
        &\leq
        M_r \log(1+\rho)
        +O(
            \log(K M_r)
        )
    \end{align}
\fi

Where $\rho \triangleq \frac{P}{N_0}$.
The last transition follows from Jensen's Inequality and the fact that $|[\mathbf{H}_k]_{i,j}|^2\sim\mathrm{Exp}(1)$.
We justify this claim with Lemma \ref{lem:CharaErr_yi2_dist} below.

For comparison, we are also interested in the settings where all users are already scheduled.
Namely, users are scheduled using Round Robin (RR).
In RR, in each time slot, \emph{exactly one} user (whose identity is well-known) transmits but may transmit at the highest rate possible.
Hence, the average ergodic Rate approximation is $\frac{K}{N}$ times the approximation for a single user given in \cite[Equation~(37)]{MIMO_MAC}:
\begin{align}
    \label{eq:ergodic_Cap_RR}
    C_{RR}
    &\approx
    \frac{K}{N}\cdot M_r \log(1+\rho)
\end{align}

\section{A MIMO-GT-Based Transmission Scheme}
\label{sect:sol_scheme}

\label{subsect:my_scheme}
In this section, we describe our suggested scheme in depth.
The scheme is comprised of three parts; codebook generation, transmission scheme, and the receiver algorithm.
The codebook generation is a random codebook whose components are i.i.d. Bernoulli random variables.
To transmit a codeword, each transmitter beamforms its signals to null the energy at all antennas \emph{whose indices correspond to zeros in the desired codeword}.
The receiver obtains the signals, and compares them to some threshold, creating a binary vector.
This binary vector is sent to a GT-decoding algorithm, namely Noisy CoMa, to obtain the codewords.
We now describe each part in depth.

First, we generate $N \mathcal{M}$ binary codewords and distribute $\mathcal{M}$ codewords to each user.
Each codeword is of length $M_t = M_r$.
Each bit in these codewords is generated using i.i.d. Bernoulli distribution with parameter $p$, which would be determined later.

To transmit the $j^{\mathrm{th}}$ codeword, $\underline{c}_j\in\{0,1\}^{M_r\times1}$, the $i^{\mathrm{th}}$ user takes the following procedure:
Let $\mathcal{Z}_j \triangleq \{l: [\underline{c}_j]_l = 0\}$.
Construct $\mathbf{H}_{\mathcal{Z}_j} \triangleq \{([\mathbf{H}_i^T]_l)^T\}_{l\in \mathcal{Z}_j}\in\mathbb{C}^{|\mathcal{Z}_j|\times M_t}$.
That is, collect all rows of $\mathbf{H}_i$ whose index corresponds to a zero in $\underline{c}_j$.
Now, calculate the orthonormal basis of $\mathbf{H}_{\mathcal{Z}_j}$'s \emph{nullspace}, and take an arbitrary linear combination of them, which holds the power constraint, to obtain $\underline{x}_i$.
% Normalize the linear combination to obtain $\underline{w}_i$.
We dub this technique as "one-dimensional Randomized Zero-Forcing" (RZF) beamforming.
The version of RZF we used in the simulations appears in algorithm \ref{alg:RZF}.
% Finally, amplify $\underline{w}_i$ to hold the power constraint and obtain $\underline{x}_i$.

\begin{algorithm}[!htbp]
    \caption{Randomized Zero-Forcing ($\mathbf{H}$, $\underline{c}$)}
    \label{alg:RZF}
    \textbf{Input:} \\
        \hspace*{\algorithmicindent} A channel matrix, $\mathbf{H}$ \\
        \hspace*{\algorithmicindent} Codeword to transmit, $\underline{c}$.\\
    \textbf{Output:} \\
        \hspace*{\algorithmicindent} Legal signal vector to transmit, $\underline{x}$ \\
    \textbf{Algorithm:} 
    \begin{algorithmic}
        \State Calculate $\mathcal{Z} \triangleq \{i: [\underline{c}]_i = 0\}$
        % \Comment{Indices of zeros in $\underline{c}$}
        \State Construct $\mathbf{H}_{\mathcal{Z}} \triangleq \{([\mathbf{H}^T]_l)^T\}_{l\in \mathcal{Z}}$
        % \in\mathbb{C}^{|\mathcal{Z}|\times M_t}$
        % \Comment{Rows correspond to zeros in $\underline{c}$}
        \State   $\mathbf{V}_{\mathcal{Z}}\gets\mathrm{Orth}(\mathrm{NullSpace}(\mathbf{H}_{\mathcal{Z}}))$
        \Comment{$\mathbf{H}_{\mathcal{Z}}$'s nullspace orthonormal basis}.
        \State
        $\underline{v} \gets \sum_{m=1}^{M_r - |\mathcal{Z}|} [\mathbf{V}_{\mathcal{Z}}]_m$ 
        \Comment{Arbitrary vector spanned by $\mathbf{V}_{\mathcal{Z}}$}
        \State $\underline{x} \gets \frac{\sqrt{P}}{\norm{\underline{v}}_2}\cdot\underline{v}$
        \Comment{Ensure $\underline{x}$ holds the power constraint}
        \State return $\underline{x}$
    \end{algorithmic}
\end{algorithm}
A user utilizing RZF, in the absence of the additive noise, assures that the receiver reads no energy from the antennas whose indices correspond to zeros.
In other antennas, there is some desirable (and optimizable) energy leakage.
$K$ users are transmitting similar signals, summed by the channel.
As a result, the receiver obtains energy at antennas corresponding to at least one '1', resulting in a Boolean sum of all transmitted codewords.
Due to the additive noise, the receiver uses an energy threshold and a relaxed decoding algorithm to obtain all $K$ codewords.
We note that RZF is not optimal;
It is possible to pick a vector from $\mathbf{H}_{\mathcal{Z}_j}$'s nullspace that maximizes the SNR at the antennas corresponding to ones.
However, analyzing the scheme with the optimized vector can be complex and does not substantially change the qualitative discussion in terms of the order-optimality we wish to accomplish.
% Additionally, in most realistic cases, $M_r$ is greater than $M_t$.
% It is possible that each user nulls its energy on the antennas corresponding to zeros and has no more degrees of freedom for optimization, resulting in the transmission of an unoptimized vector.

The receiver obtains $\underline{y}$ according to (\ref{eq:SysMod_y}), and compares $|[\underline{y}]_i|^2$ to an energy threshold $N_0 \gamma$ for all $i$.
$\gamma$ will be selected later.
The result of the comparison, $\underline{Y}$, is the result vector in the GT context.
Clearly, a hard decision using energy detection may introduce erroneous bits in $\underline{Y}$.
The errors are characterized by crossover probabilities from '1' to '0' and vice-versa, denoted by $q_{10}$ and $q_{01}$, respectively.
They are given in the following two results.
\begin{lemma}
    \label{eq:CharaErr_q01}
    For any $\gamma$, the crossover probability from '0' to '1' is
    \begin{gather}
        \nonumber
        q_{01} = e^{-\gamma}.
    \end{gather}
\end{lemma}
Note that $q_{01}$ does not depend on $N_0$, as the threshold is normalized by it.
\begin{lemma}
    \label{eq:CharaErr_q10}
    For any $\gamma$, the crossover probability from '1' to '0' is
    \begin{gather}
        \nonumber
        q_{10}
        =
        \sum_{j=1}^K \binom{K}{j}\frac{p^j(1-p)^{K-j}}{1-(1-p)^K}\bigg(1 - \exp\bigg\{-\frac{\gamma}{j \rho + 1}\bigg\}\bigg).
    \end{gather}
\end{lemma}
\noindent
The proofs for both results can be found in Subsection \ref{subsect:BER_ErrProb}.

We then use $\underline{Y}$ as an input to the Noisy CoMa algorithm, to obtain the messages.
Noisy CoMa outputs \emph{all} codewords with at least $1-q_{10}(\Delta + 1)$ common '1's with $\underline{Y}$ ($\Delta$ will be selected later).
Since it has the complete codebook, the decoder also infers the users' identities \emph{without a dedicated header}.
% The decoding step is always perfect when no additive noise is present, so the receiver may use Maximum-Likelihood (ML) decoder or (regular) CoMa as presented in \cite{SecureGroupTesting, EffiColl} to obtain vanishing error probability.
% In \cite{gtSchemeAsChannel, gtSchemeAsChannelCorrection}, Atia et al. have bounded the error probability of the ML decoder by using the union bound and dividing the error events according to the number of missing true items.
% In Noisy CoMa, on the other hand, it is possible to obtain any number of codewords between 0 and $N$, so we also have to consider the excess words.

Noisy CoMa may output any number of codewords between 0 and $N\mathcal{M}$ (consequently, any users ranging from 0 to $N$).
Hence, we have to consider two types of errors;
the first is miss-detection, where Noisy CoMa fails to find at least one transmitted codeword.
The other error is a false alarm, where Noisy CoMa declares at least one excess codeword (that was not transmitted).
The probabilities of these events are denoted by $p_{MD}$ and $p_{FA}$, respectively.
$p_{FA}$ also covers the event of identical codewords by its definition.
If $p_e$ is the error probability of MIMO-GT, then $p_e\leq p_{MD} + p_{FA}$.
Our main result is the following.
\begin{theorem}
    \label{theo:main_result}
    Fix $N$, $K$, and $\mathcal{M}$.
    Let $\delta > 0$.
    Set \mbox{$M_t=M_r \geq (1+\delta)\beta K\ln N\mathcal{M}$} for some constant $\beta \geq 1$.
    Then, MIMO-GT achieves $\max\{p_{FA}, p_{MD}\}\leq (N\mathcal{M})^{-\delta}$.
    Consequently, MIMO-GT is message-user reliable.
\end{theorem}
We note that Theorem \ref{theo:main_result} defines the relationship between $K$, $N$, $\mathcal{M}$ and $M_r$ such that the error probability is less than $(N\mathcal{M})^{-\delta}$.
As long as $M_r\geq (1+\delta)\beta K\ln N\mathcal{M}$, the theorem asserts a vanishing error probability.
The main concern of this work is to attain vanishing error probability while minimizing receiver complexity - including run-time and hardware requirements embodied in the number of antennas.
Noisy CoMa's run-time is dictated by the length of the GT result vector, which is the number of receiving antennas, $M_r$, in MIMO-GT and the product $N\mathcal{M}$.

The minimizer, $\beta^*$, is a function of $K$, $\rho$, $\gamma$, $\Delta$ and $p$ as we show in Subsection \ref{subsect:noisy_CoMa}.
If we write $p=\frac{\alpha}{K}$, a common choice in GT, and bound $\gamma$ to the interval $[1,\max\{1, \rho\}]$ then, for large enough $K$, $\beta^*$ is bounded by terms independent of $N$, $\mathcal{M}$ or $K$ for \emph{any} $\rho$.
\begin{align}
    1\leq \beta^* \leq
    \frac{8 e^{2\max\{\rho,1\}} (\rho+1)^2}{3\alpha (1-\frac{\alpha}{2})^4 \rho^2}
\end{align}
The bound is loose for \emph{high} SNR regions, but it is possible to show that $\beta^*$ converges to some constant when the SNR grows.
We elaborate on $\beta^*$'s scaling laws in Subsection \ref{subsect:beta_scale}.

Since MIMO-GT sends $K\log N\mathcal{M}$ bits per channel use, assuming the choice of the best possible $M_r$, MIMO-GT's sum-rate is
\begin{align}
    \label{eq:compare_our_rate}
    R
    &=
    K \log N\mathcal{M}
    =
    \frac{M_r}{(1+\delta)\beta^*\ln 2}.
\end{align}
Similarly, the Spectral Efficiency of MIMO-GT is
\begin{align}
    \label{eq:compare_our_spec_eff}
    \eta
    &=
    \frac{K\log N\mathcal{M}}{M_r}
    =
    \frac{1}{(1+\delta)\beta^*\ln 2}.
\end{align}
We would like to emphasize that $\beta^*$ is a function of $\rho$, hence both (\ref{eq:compare_our_rate}) and (\ref{eq:compare_our_spec_eff}).

The system's sum-rate in (\ref{eq:compare_our_rate}) can be compared with (\ref{eq:ergodic_Cap_K_CSIT}) and (\ref{eq:ergodic_Cap_RR}).
We obtain the following ratios.
\begin{align}
    \label{eq:compare_rate_ratio_CSIT_only}
    \frac{C_{Full\ CSI}}{R}
    &\approx
    (1+\delta)\beta^* \ln(1+\rho)
    +
    O\bigg(
        \frac{\log(K^2\log N\mathcal{M})}{K\log N\mathcal{M}}
    \bigg)
    \\\label{eq:compare_rate_ratio_RR}
    \frac{C_{RR}}{R}
    &\approx
    \frac{K}{N} (1+\delta)\beta^* \ln(1+\rho)
\end{align}
We observe two kinds of losses;
The first is an SNR loss, where we analyze how the rate ratios scale with $\rho$.
The other loss is the User-Codebook Loss, where we observe how the rate ratios scale with $N$, $\mathcal{M}$, and $K$.

In (\ref{eq:compare_rate_ratio_CSIT_only}), the User-Codebook Loss \emph{tends to zero}, so MIMO-GT is order-optimal when either the number of users or the number of messages grows.
The User-Codebook Loss in (\ref{eq:compare_rate_ratio_RR}) vanish when $N\to\infty$ and $K = o(N)$.
I.e., if $K$ is small enough, scheduling is ineffective as it forces all users but one to idle, and MIMO-GT is far superior.

The SNR Loss, however, scales differently.
When $\rho\to0$ $\beta^*\to\infty$ by corollary \ref{lem:beta_scale_rho} below, hence all ratios tend to infinity.
This is a direct result of Shannon's Power Efficiency Limit (SPEL); it is a lower bound on the energy per bit, equivalently on $\rho$, below a communication system cannot achieve vanishing error rates.
This law limits our system at low SNR due to the energy detection phase.
% TODO: Change me?
% In the Full CSI settings, users cooperate to create dedicated virtual channels resulting in non-zero sum-rates.
% Cooperation is not possible in MIMO-GT due to the lack of irrelevant CSI.
% Additionally, SPEL restricts the performance of energy detection in low SNRs.
% When $\rho\to\infty$, the SNR loss scales like $O(\log\rho)$.

% \label{subsect:disc}
% \input{TexFiles/05 - Noisy CoMa/Discussion}

\section{Analysis - Direct Result}
\label{sect:scheme_analysis}
This section analyzes MIMO-GT's error probability and scaling laws.
In Subsection \ref{subsect:BER_ErrProb} we calculate $q_{10}$ and $q_{01}$.
We elaborate on Noisy CoMa's performance analysis in Subsection \ref{subsect:noisy_CoMa}, and we study $\beta$'s scaling laws in Subsection \ref{subsect:beta_scale}.
In Section \ref{sect:converse}, we give a matching converse result and discuss its tightness.

\subsection{Calculating the Crossover Probabilities}
\label{subsect:BER_ErrProb}
MIMO-GT uses energy detectors to implement hard decisions at each antenna.
If no additive noise exists, the result vector, $\underline{Y}$, is a Boolean Sum of the $K$ transmitted messages.
When the additive noise is present, we use an energy threshold to convert $|\underline{y}|^2$ to $\underline{Y}$.
Therefore, we are interested in the following random variable.
\begin{align}
    \label{eq:CharaErr_Yi_def}
    [\underline{Y}]_i
    =
    \begin{cases}
        1 & |[\underline{y}]_i|^2 > N_0 \gamma \\
        0 & |[\underline{y}]_i|^2 \leq N_0 \gamma
    \end{cases}
\end{align}
$[\underline{Y}]_i$ estimates whether \emph{at least one} user targeted the $i^{\mathrm{th}}$ antenna.
$[\underline{Y}]_i$ are i.i.d due to the codebook construction and channel statistics.
Naturally, the estimation may err - either due to weak reception (caused by deep fade or strong canceling noise), denoted by $q_{10}$, or since the additive noise overcame the threshold when no user targeted the antenna (denoted by $q_{01}$).
Calculating these probabilities is similar to the derivation of a non-coherent detector for i.i.d Rayleigh fading channels in \cite[Chapter~3.1.1]{tse_book}, albeit adjusted for energy detection in a MU-MIMO environment.
Note that the detection differs from that in \cite{enrgyDetectMIMO}, since our receiver has no CSI.

To calculate $q_{01}$ and $q_{10}$, we calculate the distribution of $|[\underline{y}]_i|^2$, conditioned on the number of users targeting the $i^{\mathrm{th}}$ antenna.
The distribution is given in the following lemma.
\begin{lemma}
\label{lem:CharaErr_yi2_dist}
Assume each transmitter uses RZF.
Let $J_i$ be the number of users targeting the $i^{\mathrm{th}}$ antenna.
Then,
\begin{align}
    \nonumber
    |[\underline{y}]_i|^2 \big| J_i
    \sim
    \mathrm{Exp}\bigg(
        \frac{1}{J_i  P+N_0}
    \bigg).
\end{align}
\end{lemma}
\begin{proof}
Each transmitter uses RZF by taking an arbitrary vector that holds the power constraint from the nullspace of their channel matrix, corresponding to the zeros in their codeword.
I.e.,
\begin{align}
    y_i = \sum_{k=1}^{J_i}\sum_{j=1}^{M_r} [\mathbf{H}_k]_{i,j} [\underline{x}_k]_j
    +
    [\underline{n}]_i
\end{align}
The inner sum, 
$\sum_{j=1}^{M_r}[\mathbf{H}_k]_{i,j} [\underline{x}_k]_j$, is the standard \emph{non-complex} inner product.
In other words, a linear transformation of $([\mathbf{H}_k^T]_i)^T$.
Since $[\mathbf{H}_k]_{i,j}$ are jointly Gaussian, the inner sum is a single CGRV whose mean is zero, and its variance is $P$ \cite[Chapter 6.4.1]{leongarcia2008}.
The outer sum sums $J_i$ i.i.d RVs, resulting in a CGRV whose mean is zero and variance is $J_i  P$.
Finally, the channel adds the additive noise, $[\underline{n}]_i$, and we obtain that $[\underline{y}]_i\sim \mathcal{CN}(0, J_i P + N_0)$.

Now, we calculate the distribution of $|[\underline{y}]_i|^2$.
We are interested in the distribution of $W\triangleq|[\underline{y}]_i^2| = \Re([\underline{y}]_i)^2 + \Im([\underline{y}]_i)^2$.
The calculation has two steps - first, we calculate $Z \triangleq \sqrt{W}$'s CDF.
Then, we show that $W\sim\mathrm{Exp}(\frac{1}{2\sigma^2})$.
For simplicity, we denote $\sigma^2 \triangleq \frac{J_i  P + N_0}{2}$.
$[\underline{y}]_i\sim \mathcal{CN}(0, 2\sigma^2)$, so both its real and imaginary parts are zero-mean 
$\sigma^2$-variance Gaussians.
I.e., $\Re([\underline{y}]_i),\Im([\underline{y}]_i)\sim\mathcal{N}(0, \sigma^2)$.
\ifTCOMorTWCDraft
    \begin{align}
        F_{Z}(z)
        &=
        \mathbb{P}(Z \leq z)
        =
       \mathbb{P}\big(\sqrt{\Re([\underline{y}]_i)^2 + \Im([\underline{y}]_i)^2} \leq z\big)
        \\
        &=
        \iint\limits_{\sqrt{x^2+y^2}\leq z} \frac{1}{2\pi \sigma^2}\exp\bigg\{-\frac{x^2+y^2}{2\sigma^2}\bigg\} dx dy
        % \\
        % &
        =
        \int_{0}^z\int_{0}^{2\pi} \frac{r}{2\pi \sigma^2}\exp\bigg\{-\frac{r^2}{2\sigma^2}\bigg\} d\theta dr
        \\
        &
        =
        \int_{0}^z \frac{r}{\sigma^2}\exp\bigg\{-\frac{r^2}{2\sigma^2}\bigg\} dr
    \end{align}
\else
    \begin{align}
        F_{Z}(z)
        &=
        \mathbb{P}(Z \leq z)
        \\
        &=
       \mathbb{P}\big(\sqrt{\Re([\underline{y}]_i)^2 + \Im([\underline{y}]_i)^2} \leq z\big)
        \\
        &=
        \iint\limits_{\sqrt{x^2+y^2}\leq z} \frac{1}{2\pi \sigma^2}\exp\bigg\{-\frac{x^2+y^2}{2\sigma^2}\bigg\} dx dy
        \\
        &=
        \int_{0}^z\int_{0}^{2\pi} \frac{r}{2\pi \sigma^2}\exp\bigg\{-\frac{r^2}{2\sigma^2}\bigg\} d\theta dr
        \\
        &=
        \int_{0}^z \frac{r}{\sigma^2}\exp\bigg\{-\frac{r^2}{2\sigma^2}\bigg\} dr
    \end{align}
\fi

By deriving the CDF, we obtain that $Z$'s PDF is
\begin{align}
    \label{eq:CharaErr_Z_PDF}
    f_{Z}(z) = 
    \begin{cases}
        \frac{z}{\sigma^2}   e^{-\frac{z^2}{2\sigma^2}} & z\geq 0 \\
        0 & z < 0
    \end{cases}.
\end{align}
When $w < 0$, there is no solution to $w = z^2$ (as a function of $z$).
Hence, $f_W(w) = 0$ for any $w < 0$.
In any other case, $z = \pm\sqrt{w}$.
Now we are ready to calculate $W$'s PDF.
\ifTCOMorTWCDraft
    \begin{align}
        f_{W}(w)
        &=
        \label{eq:RVtransformTheo}
        \frac{f_{Z}(-\sqrt{w})}{|\frac{\partial w}{\partial z}|} + \frac{f_{Z}(\sqrt{w})}{|\frac{\partial w}{\partial z}|}
        =
        \frac{1}{2\sqrt{w}}\big(
            f_{Z}(-\sqrt{w}) + f_{Z}(\sqrt{w})
        \big)
        \\
        &=
        \frac{1}{2\sqrt{w}}\bigg(
            0
            +
            \frac{\sqrt{w}}{\sigma^2}   e^{-\frac{w}{2\sigma^2}}
        \bigg)
        =
        \frac{1}{2\sigma^2}   e^{-\frac{w}{2\sigma^2}}
    \end{align}
\else
    \begin{align}
        f_{W}(w)
        &=
        \label{eq:RVtransformTheo}
        \frac{f_{Z}(-\sqrt{w})}{|\frac{\partial w}{\partial z}|} + \frac{f_{Z}(\sqrt{w})}{|\frac{\partial w}{\partial z}|}
        \\
        &
        =
        \frac{1}{2\sqrt{w}}\big(
            f_{Z}(-\sqrt{w}) + f_{Z}(\sqrt{w})
        \big)
        \\
        &=
        \frac{1}{2\sqrt{w}}\bigg(
            0
            +
            \frac{\sqrt{w}}{\sigma^2}   e^{-\frac{w}{2\sigma^2}}
        \bigg)
        \\
        &=
        \frac{1}{2\sigma^2}   e^{-\frac{w}{2\sigma^2}}
    \end{align}
\fi

Step (\ref{eq:RVtransformTheo}) is the Random Variable Transformation Theorem \cite[Chapter 6]{leongarcia2008}.
Finally, $W$'s PDF is given by
\begin{align}
    \label{eq:CharaErr_W_PDF}
    f_{W}(w) = 
    \begin{cases}
        \frac{1}{2\sigma^2}   e^{-\frac{w}{2\sigma^2}} & w\geq 0 \\
        0 & w < 0
    \end{cases}.
\end{align}
Which is the PDF of an exponentially distributed random variable with mean $2\sigma^2$.
I.e., $|[\underline{y}]_i^2|\sim\mathrm{Exp}(\frac{1}{J_i  P + N_0})$.
\end{proof}

When each transmitter has a random codebook generated by i.i.d coin tosses with probability $p$ for '1', $J_i\sim \text{Bin}(K, p)$.
We can thus calculate the probabilities
\begin{align}
    \label{eq:CharaErr_q01_def}
    q_{01}
    &=
    \mathbb{P}(|[\underline{y}]_i|^2 > N_0 \gamma\ |\ J_i = 0)
    \\\label{eq:CharaErr_q10_def}
    q_{10}
    &=
    \mathbb{P}(|[\underline{y}]_i|^2\leq N_0 \gamma\ |\ J_i \geq 1).
\end{align}
We will start with $q_{01}$.
\begin{proof}[Proof of Lemma \ref{eq:CharaErr_q01}]
    \ifTCOMorTWCDraft
        \begin{align}
            q_{01}
            &=
            \mathbb{P}(|[\underline{y}]_i|^2 > N_0 \gamma\ |\ J_i = 0)
            =
            \int_{N_0 \gamma}^\infty \frac{1}{N_0}e^{-\frac{t}{N_0}} dt
            =
            e^{-\gamma}
        \end{align}
    \else
        \begin{align}
            q_{01}
            &=
            \mathbb{P}(|[\underline{y}]_i|^2\geq N_0 \gamma\ |\ J_i = 0)
            \\
            &=
            \int_{N_0 \gamma}^\infty \frac{1}{N_0}e^{-\frac{t}{N_0}} dt
            \\
            &=
            e^{-\gamma}
        \end{align}
    \fi
\end{proof}
Now, we move to calculate $q_{10}$.
\begin{proof}[Proof of Lemma \ref{eq:CharaErr_q10}]
    \ifTCOMorTWCDraft
        \begin{align}
            q_{10}
            &=
            \mathbb{P}(|[\underline{y}]_i|^2\leq N_0 \gamma\ |\ J_i \geq 1)
            =
            \frac{1}{\mathbb{P}(J_i\geq1)}
            \sum_{j=1}^K \mathbb{P}(|[\underline{y}]_i|^2\leq N_0 \gamma\ |\ J_i = j)
            \\
            &=
            \sum_{j=1}^K \binom{K}{j}\frac{p^j(1-p)^{K-j}}{1-(1-p)^K}  \mathbb{P}(|y_i|^2\leq N_0 \gamma\ |\ J_i=j)
            \\
            &=
            \sum_{j=1}^K \binom{K}{j}\frac{p^j(1-p)^{K-j}}{1-(1-p)^K}\bigg(1 - \exp\bigg\{-\frac{N_0 \gamma}{j   P + N_0}\bigg\}\bigg)
            \\
            &=
            \sum_{j=1}^K \binom{K}{j}\frac{p^j(1-p)^{K-j}}{1-(1-p)^K}\bigg(1 - \exp\bigg\{-\frac{\gamma}{j   \rho + 1}\bigg\}\bigg)   
        \end{align}
    \else
        \begin{align}
            q_{10}
            &=
            \mathbb{P}(|[\underline{y}]_i|^2\leq N_0 \gamma\ |\ J_i \geq 1)
            \\
            &=
            \frac{1}{\mathbb{P}(J_i\geq1)}
            \sum_{j=1}^K \mathbb{P}(|[\underline{y}]_i|^2\leq N_0 \gamma\ |\ J_i = j)
            \\
            &=
            \sum_{j=1}^K \binom{K}{j}\frac{p^j(1-p)^{K-j}}{1-(1-p)^K}  \mathbb{P}(|y_i|^2\leq N_0 \gamma\ |\ J_i=j)
            \\
            &=
            \sum_{j=1}^K \binom{K}{j}\frac{p^j(1-p)^{K-j}}{1-(1-p)^K}\bigg(1 - \exp\bigg\{-\frac{N_0 \gamma}{j   P + N_0}\bigg\}\bigg)
            \\
            &=
            \sum_{j=1}^K \binom{K}{j}\frac{p^j(1-p)^{K-j}}{1-(1-p)^K}\bigg(1 - \exp\bigg\{-\frac{\gamma}{j   \rho + 1}\bigg\}\bigg)   
        \end{align}
    \fi
\end{proof}
\subsection{Decoding Error Probability}
\label{subsect:noisy_CoMa}
We shall follow the footsteps of \cite[Section~V.B]{CoMa}, which analyzed the Noisy CoMa algorithm for a Binary Symmetric Channel.
However, herein, the error pattern is different, with non-symmetric errors; hence the analysis differs.
We formally define the criterion used by Noisy CoMa to obtain the messages from the output vector $\underline{Y}$.
\begin{definition}[Noisy CoMa Decision Criterion]
    \label{def:CoMa_jWordDecisionRule}
    Fix $\Delta > 0$.
    Denote $\mathrm{supp}(\underline{x})$ as the set of indices where $\underline{x}$ has non-zero components.
    Let $\mathcal{T}_j\triangleq|\mathrm{supp}(\underline{c}_j)|$ and $\mathcal{S}_j\triangleq|\mathrm{supp}(\underline{c}_j)\cap\mathrm{supp}(\underline{Y})|$.
    Noisy CoMa's declares that $\underline{c}_j$ has been transmitted if and only if $\mathcal{S}_j \geq \mathcal{T}_j(1-q_{10}(\Delta+1))$.
\end{definition}
The relaxation criterion, $1-q_{10}(\Delta + 1)$, is the same as in \cite[Chapter~IV]{robin2021capacity}.
To put it simply, Noisy CoMa examines all codewords and discards all "definitely-not-transmitted" codewords (definitely-not-defective items in the GT context) whose matching fraction with $\underline{Y}$ is less than $1-q_{10}(\Delta + 1)$.

First, we consider the probability that some antenna reads '1', $p_1$.
I.e., $p_1 \triangleq \mathbb{P}([\underline{Y}]_i = 1)$.
We can calculate this probability by calculating its complement, 
\begin{align}
    p_0 = (1-(1-p)^K)q_{10} + (1-p)^K (1-q_{01}).
\end{align}
Intuitively, $p_1$ is the probability that some component of a transmitted codeword is "hidden" by other codewords or noise.

\begin{lemma}
    \label{theo:CoMa_pMD_to0_suff_cond}
    Fix some $\delta > 0$.
    Set $M_r \geq (1+\delta)\beta_1 K\ln N\mathcal{M}$.
    If 
    \begin{align}
        \nonumber
        \beta_1 \geq \frac{1}{K p\big(1-\exp\big\{-2(q_{10}\Delta)^2\big\}\big)}
    \end{align}
    then $p_{MD}\leq (N\mathcal{M})^{-\delta}$.
\end{lemma}
\begin{proof}
    To prove lemma \ref{theo:CoMa_pMD_to0_suff_cond}, we introduce the following proposition.
    \begin{proposition}
        \label{theo:CoMa_pMD_ub}
        The probability of missing at least one true codeword is bounded from above by
        \begin{align}
            \nonumber
            p_{MD} \leq K   \exp\bigg\{-M_r   p   \big(1-e^{-2(q_{10}\Delta)^2}\big)\bigg\}.
        \end{align}
    \end{proposition}
    \begin{proof}
        Denote $F_{\mathrm{Bin}}(k; n, p)\triangleq \sum_{i=0}^{k}\binom{n}{i}p^i(1-p)^{n-i}$.
        \ifTCOMorTWCDraft
            \begin{align}
                \label{eq:pMD:upperBound:proof:A}
                p_{MD}
                &\leq
                K \sum_{i=0}^{M_r} \mathbb{P}(\mathcal{T}_1 = i) \mathbb{P}(\mathcal{S}_1 < \mathcal{T}_1 (1-q_{10}(\Delta+1))|\mathcal{T}_1 = i)
                \\\label{eq:pMD:upperBound:proof:B}
                &=
                K  \sum_{i=0}^{M_r} \binom{M_r}{i}p^i(1-p)^{M_r-i}
                \mathbb{P}(\mathcal{S}_1 < i(1-q_{10}(\Delta+1)))
                \\
                \label{eq:pMD:upperBound:proof:C}
                &=
                K  \sum_{i=0}^{M_r} \binom{M_r}{i}p^i(1-p)^{M_r-i}
                \sum_{l=i-i(1-q_{10}(\Delta+1))}^i \binom{i}{l} q_{10}^i (1-q_{10})^{i-l}
                \\\label{eq:pMD:upperBound:proof:D}
                &\leq
                K \sum_{i=0}^{M_r} \binom{M_r}{i}p^i(1-p)^{M_r-i} e^{-2i(q_{10}\Delta)^2}
                \\\label{eq:pMD:upperBound:proof:E}
                &=
                K \big(1-p\big(1- e^{-2(q_{10}\Delta)^2}\big)\big)^{M_r}
                \\\label{eq:pMD:upperBound:proof:F}
                &\leq
                K   \exp\bigg\{-M_r   p  \big(1-e^{-2(q_{10}\Delta)^2}\big)\bigg\}
            \end{align}
        \else
            \begin{align}
                \nonumber
                p&_{MD}
                \leq
                \\\label{eq:pMD:upperBound:proof:A}
                &
                K \sum_{i=0}^{M_r} \mathbb{P}(\mathcal{T}_1 = i)   \mathbb{P}(\mathcal{S}_1 < \mathcal{T}_1 (1-q_{10}(\Delta+1))|\mathcal{T}_1 = i)
                \\\label{eq:pMD:upperBound:proof:B}
                &=
                K  \sum_{i=0}^{M_r} \binom{M_r}{i}p^i(1-p)^{M_r-i}
                \mathbb{P}(\mathcal{S}_1 < i(1-q_{10}(\Delta+1)))
                \\
                \label{eq:pMD:upperBound:proof:C}
                &=
                K  \sum_{i=0}^{M_r} \binom{M_r}{i}p^i(1-p)^{M_r-i}
                \\\nonumber
                &\qquad
                \sum_{l=i-i(1-q_{10}(\Delta+1))}^i \binom{i}{l} q_{10}^i (1-q_{10})^{i-l}
                \end{align}
                \begin{align}
                \label{eq:pMD:upperBound:proof:D}
                &\leq
                K \sum_{i=0}^{M_r} \binom{M_r}{i}p^i(1-p)^{M_r-i} e^{-2i(q_{10}\Delta)^2}
                \\\label{eq:pMD:upperBound:proof:E}
                &=
                K \big(1-p\big(1- e^{-2(q_{10}\Delta)^2}\big)\big)^{M_r}
                \\\label{eq:pMD:upperBound:proof:F}
                &\leq
                K   \exp\bigg\{-M_r   p  \big(1-e^{-2(q_{10}\Delta)^2}\big)\bigg\}
            \end{align}
        \fi
        In (\ref{eq:pMD:upperBound:proof:A}), we used the union bound and the law of total probability.
        (\ref{eq:pMD:upperBound:proof:B}) is derived from the random codebook construction where $\mathcal{S}_1\sim\text{Bin}(M_r, p)$.
        In (\ref{eq:pMD:upperBound:proof:C}), we used the fact that the local decision rule is identical among the antennas, so the probability for bit flips is symmetric (the number of flipped bits is binomially distributed with parameters $M_r$, $q_{10}$).
        Additionally, we used the binomial distribution's symmetry.
        I.e., $F_{\mathrm{Bin}}(k; n, p) = F_{\mathrm{Bin}}(n-k; n, 1-p)$.
        In equation (\ref{eq:pMD:upperBound:proof:D}), we used Hoeffding bound, $F_{\mathrm{Bin}}(k; n, p) \leq \exp\{-2n(p-\frac{k}{n})^2\}$ for any $p-\frac{k}{n} > 0$ \cite{vershynin2018HighDimProb}.
        (\ref{eq:pMD:upperBound:proof:E}) used the binomial theorem to combine the sum of products into a power of a sum.
        In (\ref{eq:pMD:upperBound:proof:F}), we used the Taylor Expansion of $\exp\{-x(1-a)\}$ at $x_0 = 0$.
        That is, $e^{-x(1-a)} = 1-(1-a)x + O((1-a)^2x^2)$.
    \end{proof}
    Now we are ready to prove lemma \ref{theo:CoMa_pMD_to0_suff_cond}.
    \ifTCOMorTWCDraft
        \begin{align}
            p_{MD}
            &\leq
            K\exp\bigg\{-M_r   p   \big(1-e^{-2(q_{10}\Delta)^2}\big)\bigg\}
            \\\label{eq:pMD:upperBound:suffCond:proof:A}
            &=
            K\exp\bigg\{-(1+\delta)\beta_1 (K\ln N\mathcal{M})p\big(1-e^{-2(q_{10}\Delta)^2}\big)\bigg\}
            \\\label{eq:pMD:upperBound:suffCond:proof:B}
            &\leq
            K \exp\bigg\{-\frac{(1+\delta)(K\ln N\mathcal{M})p \big(1-e^{-2(q_{10}\Delta)^2}\big)}{K p\big(1-e^{-2(q_{10}\Delta)^2}\big)} \bigg\}
            \\
            &=
            K \exp\big\{-(1+\delta) \ln N\mathcal{M} \big\}    
            \\\label{eq:K_scaling_Law_proof_pMD}
            &= K (N\mathcal{M})^{-(1+\delta)}
            \leq (N\mathcal{M}) (N\mathcal{M})^{-(1+\delta)} = (N\mathcal{M})^{-\delta}
        \end{align}
    \else
        \begin{align}
            p&_{MD}
            \leq
            K\exp\bigg\{-M_r   p   \big(1-e^{-2(q_{10}\Delta)^2}\big)\bigg\}
            \\\label{eq:pMD:upperBound:suffCond:proof:A}
            &=
            K\exp\bigg\{-(1+\delta)\beta_1 (K\ln N\mathcal{M})p\big(1-e^{-2(q_{10}\Delta)^2}\big)\bigg\}
            \\\label{eq:pMD:upperBound:suffCond:proof:B}
            &\leq
            K \exp\bigg\{-\frac{(1+\delta)(K\ln N\mathcal{M})p \big(1-e^{-2(q_{10}\Delta)^2}\big)}{K p\big(1-e^{-2(q_{10}\Delta)^2}\big)} \bigg\}
            \\
            &=
            K \exp\big\{-(1+\delta) \ln N\mathcal{M} \big\}    
            \\\label{eq:K_scaling_Law_proof_pMD}
            &= K (N\mathcal{M})^{-(1+\delta)}
            \leq (N\mathcal{M}) (N\mathcal{M})^{-(1+\delta)} = (N\mathcal{M})^{-\delta}
        \end{align}
    \fi
    In (\ref{eq:pMD:upperBound:suffCond:proof:A}) and (\ref{eq:pMD:upperBound:suffCond:proof:B}), we utilized lemma \ref{theo:CoMa_pMD_to0_suff_cond}'s conditions.
\end{proof}
\noindent
Likewise, we have a sufficient condition on $p_{FA}$.
\begin{lemma}
    \label{theo:CoMa_pFA_to0_suff_cond}
    Fix some $\delta > 0$.
    Set $M_r \geq (1+\delta)\beta_2 K\ln N\mathcal{M}$ and $\Delta < \frac{p_0}{q_{10}}-1$.
    If 
    \begin{align}
        \nonumber
        \beta_2\geq \frac{1}{K p\big(1-\exp\big\{-2(p_0-q_{10}(\Delta+1))^2\big\}\big)}
    \end{align}
    then $p_{FA}\leq (N\mathcal{M})^{-\delta}$.
\end{lemma}
\begin{proof}
    The proof of lemma \ref{theo:CoMa_pFA_to0_suff_cond} is similar to lemma \ref{theo:CoMa_pMD_to0_suff_cond}'s, with the following upper bound:
    \begin{proposition}
    \label{theo:CoMa_pFA_ub}
    Assume $\Delta < \frac{p_0}{q_{10}}-1$.
    The probability of declaring at least one false codeword is bounded from above by
    \begin{align}
        \nonumber
        p_{FA} \leq (N\mathcal{M}-K)
        \exp\bigg\{-M_r   p   \big(1-e^{-2(p_0-q_{10}(\Delta+1))^2}\big)\bigg\}.
    \end{align}
    \end{proposition}
    The additional condition $\Delta < \frac{p_0}{q_{10}}-1$ is required for the Hoeffding Bound to hold.
    \begin{proof}
    \ifTCOMorTWCDraft
        \begin{align}
            p_{FA}
            &\leq
            (N\mathcal{M}-K) \sum_{i=0}^{M_r} \mathbb{P}(\mathcal{T}_1 = i) \mathbb{P}(\mathcal{S}_1 \geq \mathcal{T}_1 (1-q_{10}(\Delta+1))|\mathcal{T}_1 = i)
            \\
            &=
            (N\mathcal{M}-K)\sum_{i=0}^{M_r} \binom{M_r}{i}p^i(1-p)^{M_r-i}\sum_{l=i(1-q_{10}(\Delta+1))}^i \binom{i}{l} p_1^i (1-p_1)^{i-l}
            \\\label{eq:pFA:upperBound:proof:A}
            &\leq
            (N\mathcal{M}-K)  \sum_{i=0}^{M_r} \binom{M_r}{i}p^i(1-p)^{M_r-i} e^{-2i(1-p_1-q_{10}(\Delta+1))^2}
            \\
            &=
            (N\mathcal{M}-K)  \big(1-p\big(1-e^{-2(p_0-q_{10}(\Delta+1))^2}\big)\big)^{M_r}
            \\
            &\leq
            (N\mathcal{M}-K) 
            \exp\bigg\{-M_r  p  \big(1-e^{-2(p_0-q_{10}(\Delta+1))^2}\big)\bigg\}
        \end{align}
    \else
        \begin{align}
            \nonumber
            p&_{FA}
            \leq
            (N\mathcal{M}-K) \sum_{i=0}^{M_r} \mathbb{P}(\mathcal{T}_1 = i)
            \\
            &\qquad
            \cdot\mathbb{P}(\mathcal{S}_1 \geq \mathcal{T}_1 (1-q_{10}(\Delta+1))|\mathcal{T}_1 = i)
            \\\nonumber
            &=
            (N\mathcal{M}-K)\sum_{i=0}^{M_r} \binom{M_r}{i}p^i(1-p)^{M_r-i}
            \\
            &\qquad 
            \cdot\sum_{l=i(1-q_{10}(\Delta+1))}^i \binom{i}{l} p_1^i (1-p_1)^{i-l}
            \\\nonumber
            &\leq
            (N\mathcal{M}-K)  \sum_{i=0}^{M_r} \binom{M_r}{i}p^i(1-p)^{M_r-i}
            \\\label{eq:pFA:upperBound:proof:A}
            &\qquad
            \cdot e^{-2i(1-p_1-q_{10}(\Delta+1))^2}
            % \\\nonumber
            % &=
            % (N\mathcal{M}-K)  \sum_{i=0}^{M_r} \binom{M_r}{i}p^i(1-p)^{M_r-i}\cdot
            % \\&\qquad
            % \cdot e^{-2i(p_0-q_{10}(\Delta+1))^2}
            \\
            &=
            (N\mathcal{M}-K)  \big(1-p\big(1-e^{-2(p_0-q_{10}(\Delta+1))^2}\big)\big)^{M_r}
            \\
            &\leq
            (N\mathcal{M}-K) 
            \exp\bigg\{-M_r  p  \big(1-e^{-2(p_0-q_{10}(\Delta+1))^2}\big)\bigg\}
        \end{align}
    \fi
    In equation (\ref{eq:pFA:upperBound:proof:A}) we used Hoeffding bound like in the proof of proposition \ref{theo:CoMa_pMD_ub}, which is applicable when $\Delta < \frac{p_0}{q_{10}}-1$.
    Any other transition is identical to the proof of proposition \ref{theo:CoMa_pMD_ub}.
    \end{proof}
    Now we prove lemma \ref{theo:CoMa_pFA_to0_suff_cond} similarly to lemma \ref{theo:CoMa_pMD_to0_suff_cond}.
    \ifTCOMorTWCDraft
        \begin{align}
            p_{FA}
            &\leq
            (N\mathcal{M}-K)
            \exp\bigg\{-M_r p \big(1-e^{-2(p_0-q_{10}(\Delta+1))^2}\big)\bigg\}
            \\\label{eq:pFA:upperBound:suffCond:proof:A}
            &=
            (N\mathcal{M}-K) \exp\bigg\{-(1+\delta)\beta_2 (K\ln N\mathcal{M})p\big(1-e^{-2(p_0-q_{10}(\Delta+1))^2}\big)\bigg\}
            \\\label{eq:pFA:upperBound:suffCond:proof:B}
            &\leq
            (N\mathcal{M}-K)\exp\bigg\{\frac{-(1+\delta)(K\ln N\mathcal{M})p}{Kp\big(1-e^{-2(p_0-q_{10}(\Delta+1))^2}\big)}\big(1-e^{-2(p_0-q_{10}(\Delta+1))^2}\big)\bigg\}
            \\\label{eq:K_scaling_Law_proof_pFA}
            &=
            (N\mathcal{M}-K)\exp\big\{-(1+\delta)\ln N\mathcal{M}\big\}
            \leq (N\mathcal{M})^{-\delta}
        \end{align}
    \else
        \begin{align}
            p_{FA}
            &\leq
            (N\mathcal{M}-K)
            \exp\bigg\{-M_r p \big(1-e^{-2(p_0-q_{10}(\Delta+1))^2}\big)\bigg\}
            \\\label{eq:pFA:upperBound:suffCond:proof:A}
            &=
            (N\mathcal{M}-K) \exp\bigg\{-(1+\delta)\beta_2 (K\ln N\mathcal{M})p
            \\\nonumber
            &\qquad
            \cdot\big(1-e^{-2(p_0-q_{10}(\Delta+1))^2}\big)\bigg\}
            \\\label{eq:pFA:upperBound:suffCond:proof:B}
            &\leq
            (N\mathcal{M}-K)\exp\bigg\{\frac{-(1+\delta)(K\ln N\mathcal{M})p}{Kp\big(1-e^{-2(p_0-q_{10}(\Delta+1))^2}\big)}
            \\\nonumber
            &\qquad
            \cdot\big(1-e^{-2(p_0-q_{10}(\Delta+1))^2}\big)\bigg\}
            \\\label{eq:K_scaling_Law_proof_pFA}
            &=
            (N\mathcal{M}-K)\exp\big\{-(1+\delta)\ln N\mathcal{M}\big\}
            \leq (N\mathcal{M})^{-\delta}
        \end{align}
    \fi
    % In both (\ref{eq:pFA:upperBound:suffCond:proof:A}) and (\ref{eq:pFA:upperBound:suffCond:proof:B}) we used lemma \ref{theo:CoMa_pFA_to0_suff_cond}'s conditions.
\end{proof}

Both propositions \ref{theo:CoMa_pMD_ub} and \ref{theo:CoMa_pFA_ub} hold due to $K\leq N$.
By carefully examining equations (\ref{eq:K_scaling_Law_proof_pMD}) and (\ref{eq:K_scaling_Law_proof_pFA}) we obtain the following corollary.
\begin{corollary}[$K$'s Scaling Law]
\ifTCOMorTWCDraft
    If $K=N^{\varepsilon}$ for some $0\leq\varepsilon\leq1$ and \break $M_r\geq (1+\delta)\max\{\beta_1,\beta_2\}K\ln N\mathcal{M}$, then $\max\{p_{MD}, p_{FA}\}\leq (N \mathcal{M})^{-\delta}$.
\else
    If $K=N^{\varepsilon}$ for some $0\leq\varepsilon\leq1$ and $M_r\geq (1+\delta)\max\{\beta_1,\beta_2\} K\ln N\mathcal{M}$, then $\max\{p_{MD}, p_{FA}\}\leq (N \mathcal{M})^{-\delta}$.
\fi
\end{corollary}

We are interested in a vanishing error probability for our scheme, so we would like to take $\beta$ such that both $p_{MD}$ and $p_{FA}$ tend to zero.
In other words, an appropriate choice of $\beta$ is the solution to the optimization problem of minimizing the number of antennas required for MIMO-GT.
\begin{align}
    \label{eq:CoMa_b_opt_prob_orig}
    \begin{aligned}
        \min_{\Delta, p,\gamma \in \mathbb{R}}  & \max\{\beta_1,
        \beta_2\}
        \\
        \textrm{s.t.} \quad & 
        % 0 < \Delta < \frac{p_0}{q_{10}}-1;\quad
        %     0 < \alpha \leq \frac{K}{2};\quad
        %     0 < \gamma
        \begin{cases}
            0 < \Delta < \frac{p_0}{q_{10}}-1 \\
            0 < p \leq \frac{1}{2}\\
            0 < \gamma 
        \end{cases}
    \end{aligned}
\end{align}

The following lemma simplifies the optimization problem.
\begin{lemma}
    \label{theo:CoMa_b_opt_prob_equiv}
    The optimization problem in (\ref{eq:CoMa_b_opt_prob_orig}) is equivalent to, and has the same solution as the following optimization problem:
    \begin{align}
        \nonumber
        \begin{aligned}
            \min_{p,\gamma \in \mathbb{R}}  & \frac{1}{K p \big(1-\exp\{-\frac{1}{2}(1-p)^{2K}(1-q_{10}-q_{01})^2\}\big)}
            \\
            \textrm{s.t.} \quad & 
            \begin{cases}
                0 \leq p \leq \frac{1}{2}\\
                0 \leq \gamma
            \end{cases}
        \end{aligned}
    \end{align}
    which has a unique solution.
\end{lemma}
The proof is technical and appears in Appendix \ref{appendix:proof_CoMa_b_opt_prob_equiv}.
% the full version.
\emph{Proof Sketch}:
The proof has five steps;
(1) simplify $\Delta$'s upper bound, $\frac{p_0}{q_{10}}-1$, to $(1-p)^K\frac{1-q_{01}-q_{10}}{q_{10}}$.
(2) eliminate the dependency on $\Delta$ and convert the minimax problem into a minimization problem (by noticing that $\beta_1$ and $\beta_2$ have opposing trends in $\Delta$, so $\Delta^*$ is their equalizer).
(3) by defining $p\triangleq \frac{\alpha}{K}$, we bound $\alpha^*$ to the interval $[0, e-1]$.
(4) show that for each $\alpha$ there exists a unique $\gamma^*$ minimizing the objective function.
By putting all steps together, by the continuity of the objective function, a solution exists.
Step (4) assures its uniqueness.
Finally, the final step shows that numerical algorithms will converge to $(p^*, \gamma^*)$, which is an interior point in $[0, \frac{1}{2}]\times [0,\infty)$, by putting all previous steps together and the fact that some boundary points are infeasible.
\subsection{Antenna Scaling Laws}
\label{subsect:beta_scale}
This subsection shows different scaling laws on $\beta^*$, the solution to the optimization problem in lemma \ref{theo:CoMa_b_opt_prob_equiv}, as a function of $K$ and $\rho$.
First, we shall show that $\beta^*$ converges to some constant term when $K\to\infty$.
\begin{proposition}
    \label{lem:betaStar:bound}
    Let $p=\frac{\alpha}{K}$, where $\alpha > 0$ is some constant.
    If $1\leq \gamma\leq \max\{1, \rho\}$, then
    \begin{align}
        \nonumber
        \beta^*
        \leq
        \frac{8 e^{2\max\{1, \rho\}} (\rho+1)^2}{3\alpha (1-\frac{\alpha}{2})^4 \rho^2}.
    \end{align}
\end{proposition}
The proof is attached in appendix \ref{appendix:proof:betaStar:bound}.
\emph{Proof Sketch}: The proofs consist of three steps; bound $q_{10}$ from above by its largest addend.
Next, bound $(1-\frac{\alpha}{K})^{2K}$ from below by substituting $K=2$.
Finally, apply Taylor series expansion successively to obtain the bound.

$\beta^*\geq 1$ and proposition \ref{lem:betaStar:bound} bounds $\beta^*$ from above.
Hence, $\beta^*$ converges to some constant when $K\to\infty$ regardless of what trends $\beta^*$ exhibits when $K$ grows.
\begin{lemma}
    \label{lem:beta_scale_K_inf}
    Let $p = \frac{\alpha}{K}$ for some $\alpha > 0$.
    Let $\beta^*$ be the solution to the optimization problem in lemma \ref{theo:CoMa_b_opt_prob_equiv}.
    Then, $\beta^*$ converges to some constant when $K\to\infty$.
\end{lemma}
\emph{Proof Sketch}:
When $p = \frac{\alpha}{K}$ we can invoke the Poisson Limit Theorem on $q_{10}$.
Any other term dependant on $K$ tends to some exponent powered by $\alpha$, so $\beta^*$ tends to some constant.
Complete proof can be found in appendix \ref{appendix:proof_beta_scale_K_inf}.

For $\beta^*$'s scaling with $\rho$, we observe that the expression in lemma \ref{theo:CoMa_b_opt_prob_equiv} is not a function of $\rho$, except for $q_{10}$ (lemma \ref{eq:CharaErr_q10}).
From this observation, we have the following (immediate) result:
\begin{corollary}
    \label{lem:beta_scale_rho}
    If $\rho\to0$ then $\beta^*\to\infty$.
    If $\rho\to\infty$ then $\beta^*\to\mathrm{const.}$
\end{corollary}

\section{Converse}
\label{sect:converse}
In this section, we shall derive a lower bound on the number of antennas.

\begin{theorem}
    \label{eq:ConverseMrBound}
    Assume $N$ users, each with a $\mathcal{M}$-sized codebook.
    Assume $K$ of them wish to transmit a single codeword.
    Then, a lower bound on the number of antennas, $M_r$, is required to obtain the $K$ messages reliably when using a hard-decision criterion at the receiver is
    \begin{align}
        \nonumber
        M_r \geq \frac{K\log \frac{N \mathcal{M}}{K}}{C_{BAC} (q_{01},q_{10})}.
    \end{align}
    Where $C_{BAC} (q_{01},q_{10})$ is the BAC capacity (\cite{BACCapacity})
    \ifTCOMorTWCDraft
        \begin{align}
            \label{eq:CharaErrCapBAC}
            &C_{BAC} (q_{01},q_{10})
            =
            \frac{q_{01}}{1-q_{01}-q_{10}} H(q_{10})
            -\frac{1-q_{10}}{1-q_{01}-q_{10}}  H(q_{01})
            +\log\bigg( 1+2^{\frac{H(q_{01})-H(q_{10})}{1-q_{01}-q_{10})}} \bigg).
        \end{align}
    \else
        \begin{align}
            \label{eq:CharaErrCapBAC}
            &C_{BAC} (q_{01},q_{10})
            =
            \frac{q_{01}}{1-q_{01}-q_{10}} H(q_{10})
            \\\nonumber
            &\qquad-\frac{1-q_{10}}{1-q_{01}-q_{10}}  H(q_{01})
            +\log\bigg( 1+2^{\frac{H(q_{01})-H(q_{10})}{1-q_{01}-q_{10})}} \bigg).
        \end{align}
    \fi
    $H(\cdot)$ is the binary entropy function.
\end{theorem}
\emph{Proof Sketch}:
The proof is similar to \cite[Chapter~IV]{CoMa2}.
Notice that the $K$ messages, their corresponding codewords, the "clean" Boolean sum, the noisy output vector, the estimated codewords, and estimated messages form a Markov Chain.
We use Fano's Inequality and Data Processing Inequality to bound the entropy of the messages and use algebraic manipulations to obtain the result.
\begin{proof}
    Denote $\underline{\tilde{Y}}$ as the noiseless $\underline{Y}$.
    It is the column-wise Boolean sum of the $K$ transmitted codewords.
    Let $\mathcal{W}\in\{1,2,\dots,\binom{N}{K}\mathcal{M}^K\}$ be the index of the set corresponding to the $K$ messages transmitted.
    The codewords corresponding to $\mathcal{W}$ are $\mathcal{C}(\mathcal{W})$.
    Observe that we can treat $\underline{Y}$ as an output of some discrete channel (characterized by $q_{10}$ and $q_{01}$) whose input is $\underline{\tilde{Y}}$.
    Notice the following Markov Chain
    \begin{align}
        \mathcal{W}\rightarrow\mathcal{C}(\mathcal{W})\rightarrow\underline{\tilde{Y}}\rightarrow \underline{Y}\rightarrow\mathcal{\hat{C}}(\hat{\mathcal{W}})\rightarrow \mathcal{\hat{W}}.
    \end{align}
    Using the definition of mutual information, we have
    \begin{align}
        \label{eq:ConverseEntropyB4Fano}
        H(\mathcal{W}) = H(\mathcal{W}|\hat{\mathcal{W}}) + I(\mathcal{W}; \mathcal{\hat{\mathcal{W}}}).
    \end{align}
    We assume we have no prior information on the messages or users, so $\mathcal{W}$ is uniform on $\{1,2,\dots \binom{N}{K}\mathcal{M}^K\}$.
    This is equivalent to $H(\mathcal{W}) = \log\binom{N}{K}\mathcal{M}^K$.
    Using Fano's Inequality, we have
    \begin{align}
        \label{eq:ConverseFano}
        H(\mathcal{W}|\mathcal{\hat{W}}) \leq 1+p_e \log\binom{N}{K}\mathcal{M}^K.
    \end{align}
    We have the following chain of inequalities
    \ifTCOMorTWCDraft
        \begin{align}
            \label{eq:Converse:Proof:A}
            I(\mathcal{W}; \mathcal{\hat{\mathcal{W}}})
            &\leq
            I(\underline{\tilde{Y}}; \underline{Y})
            =
            H(\underline{Y}) - H(\underline{Y}|\underline{\tilde{Y}})
            \\\label{eq:Converse:Proof:B}
            &=
            H(\underline{Y}) - \sum_{m=1}^{M_r} H([\underline{Y}]_m|[\underline{Y}]_1,\dots,[\underline{Y}]_{m-1},\underline{\tilde{Y}})
            \\\label{eq:Converse:Proof:C}
            &=
            \sum_{m=1}^{M_r} H([\underline{Y}]_m) - \sum_{m=1}^{M_r} H([\underline{Y}]_m|[\underline{\tilde{Y}}]_m)
            \\
            &=
            \sum_{m=1}^{M_r} \big[ H([\underline{Y}]_m) - H([\underline{Y}]_m|[\underline{\tilde{Y}}]_m) \big]
            \\
            &=
            \sum_{m=1}^{M_r} I([\underline{\tilde{Y}}]_m; [\underline{Y}]_m)
            \leq\label{eq:ConverseMutInfoUpBound}
            M_r C_{BAC} (q_{01},q_{10})
        \end{align}
    \else
        \begin{align}
            \label{eq:Converse:Proof:A}
            I(\mathcal{W}; \mathcal{\hat{\mathcal{W}}})
            &\leq
            I(\underline{\tilde{Y}}; \underline{Y})
            \\
            &=
            H(\underline{Y}) - H(\underline{Y}|\underline{\tilde{Y}})
            \\\label{eq:Converse:Proof:B}
            &=
            H(\underline{Y}) - \sum_{m=1}^{M_r} H([\underline{Y}]_m|[\underline{Y}]_1,\dots,[\underline{Y}]_{m-1},\underline{\tilde{Y}})
            \\\label{eq:Converse:Proof:C}
            &=
            \sum_{m=1}^{M_r} H([\underline{Y}]_m) - \sum_{m=1}^{M_r} H([\underline{Y}]_m|[\underline{\tilde{Y}}]_m)
            \\
            &=
            \sum_{m=1}^{M_r} \big[ H([\underline{Y}]_m) - H([\underline{Y}]_m|[\underline{\tilde{Y}}]_m) \big]
            \\
            &=
            \sum_{m=1}^{M_r} I([\underline{\tilde{Y}}]_m; [\underline{Y}]_m)
            \\\label{eq:ConverseMutInfoUpBound}
            &\leq 
            M_r C_{BAC} (q_{01},q_{10})
        \end{align}
    \fi
    (\ref{eq:Converse:Proof:A}) is valid due to the Data Processing Inequality.
    In (\ref{eq:Converse:Proof:B}), we have used the Entropy Chain Rule, whereas in (\ref{eq:Converse:Proof:C}), we used the fact that $[\underline{\tilde{Y}}]_m$ ($[\underline{Y}]_m$) is independent of $[\underline{\tilde{Y}}]_k$ ($[\underline{Y}]_k$) for all $m\neq k$.
    Now, we can put (\ref{eq:ConverseMutInfoUpBound}) and (\ref{eq:ConverseFano}) into (\ref{eq:ConverseEntropyB4Fano}) to obtain the following
    \ifTCOMorTWCDraft
        \begin{align}
            \log \binom{N}{K}\mathcal{M}^K
            &\leq
            1 + p_e \log \binom{N}{K}\mathcal{M}^K + M_r C_{BAC} (q_{01},q_{10}).
        \end{align}
    \else
        \begin{align}
            \nonumber
            \log \binom{N}{K}\mathcal{M}^K
            &\leq
            1 + p_e \log \binom{N}{K}\mathcal{M}^K
            \\
            &
            \qquad+ M_r C_{BAC} (q_{01},q_{10}).
        \end{align}
    \fi
    Rearranging both sides results in a lower bound on $M_r$
    \begin{align}
        M_r \geq \frac{(1-p_e) \log \binom{N}{K}\mathcal{M}^K-1}{C_{BAC} (q_{01},q_{10})}.
    \end{align}
    Next, using Stirling's Approximation, we have
    \ifTCOMorTWCDraft
        \begin{align}
            \log \binom{N}{K}\mathcal{M}^K
            &=
            \log \binom{N}{K}
            +
            \log \mathcal{M}^K
            \approx
            K\log \frac{N}{K}
            +
            K\log \mathcal{M}
            =
            K\log \frac{N\mathcal{M}}{K}.
        \end{align}
    \else
        \begin{align}
            \log \binom{N}{K}\mathcal{M}^K
            &=
            \log \binom{N}{K}
            +
            \log \mathcal{M}^K
            \\
            &\approx
            K\log \frac{N}{K}
            +
            K\log \mathcal{M}
            \\
            &=
            K\log \frac{N\mathcal{M}}{K} 
        \end{align}
    \fi
    Which results in
    \begin{align}
        M_r \geq \frac{(1-p_e) K\log \frac{N\mathcal{M}}{K}}{C_{BAC} (q_{01},q_{10})}.
    \end{align}
\end{proof}
Dividing the converse bound with our scheme's number of antennas yields
\ifTCOMorTWCDraft
    \begin{align}
        \frac{K\log \frac{N\mathcal{M}}{K}}{C_{BAC} (q_{01},q_{10})}
        \cdot
        \frac{1}{(1+\delta)\beta^* K\ln N\mathcal{M}}
        =
        \frac{1-\frac{\log K}{\log N\mathcal{M}}}{(1+\delta)\beta^* C_{BAC} (q_{01},q_{10})\ln 2}
    \end{align}
\else
    \begin{align}
        \nonumber
        &\frac{K\log \frac{N\mathcal{M}}{K}}{C_{BAC} (q_{01},q_{10})}
        \cdot
        \frac{1}{(1+\delta)\beta^* K\ln N\mathcal{M}}
        \\&\qquad
        =
        \frac{1-\frac{\log K}{\log N\mathcal{M}}}{(1+\delta)\beta^* C_{BAC} (q_{01},q_{10})\ln 2}
    \end{align}
\fi
which tends to a constant when $N\mathcal{M}\to\infty$.
In other words, our scheme's number of antennas is asymptotically tight when $N$ and $\mathcal{M}$ grow as long as $K < N\mathcal{M}$.
According to corollary \ref{lem:beta_scale_rho}, the same claims hold when $\rho\to\infty$ and $N$, $\mathcal{M}$ and $K$ are constant.

\section{Numerical Results}
\label{sect:sim_num_res}
In this section, we present simulation and numerical evaluation results.
The simulation had a network of $N=100$ sensors with $\mathcal{M}=1000$ codewords each.
$\rho$ is taken to be $24_{\mathrm{dB}}$\footnote{SNR required for MCS3 in 802.11ac, \cite{WiFi_Guide}}, $N_0 = 2$ and $\delta=0.33$.
We assumed $K=\log_{10} N\mathcal{M} = 5$ users wish to transmit simultaneously and solved the optimization problem in lemma \ref{theo:CoMa_b_opt_prob_equiv} to obtain $\gamma^* = 5.6583$, $p^*=0.0986$ and $\beta^*=9.048$.
The initial number of antennas at the receiver is the Converse bound in Theorem \ref{eq:ConverseMrBound}.

\begin{figure}
    \centering
    \input{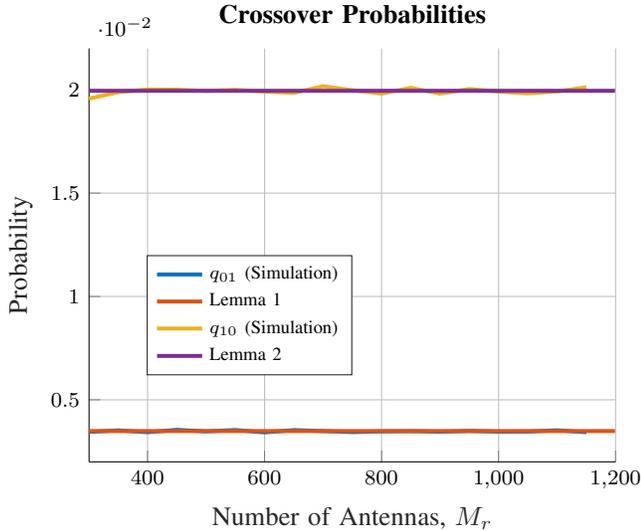}
    \caption[The detection error probabilities per antenna]{
        The detection error per antenna.
        Either an antenna fails to identify a transmission ($q_{10}$) or detects a non-existing transmission ($q_{01}$) for $N=100$, $\mathcal{M}=1000$, $K=5$, $\rho=24_{\mathrm{dB}}$, $\delta=0.33$, $\gamma^* = 11.318$ and $p^*=0.0986$.
    }
    \label{fig:res_q10_q01}
\end{figure}

\begin{figure}
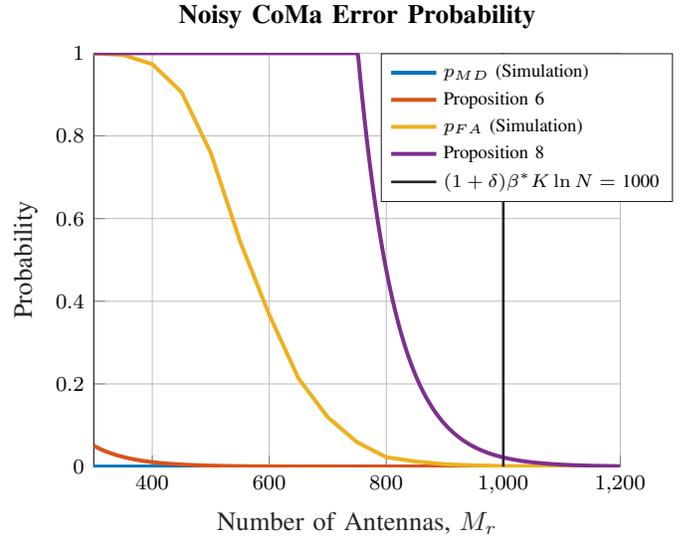

    \centering
    \ifTCOMorTWCDraft
        \input{TexFiles/TexImages/pFA_pMD_TCOM_TWC_draft.tex}
    \else
        \input{TexFiles/TexImages/pFA_pMD.tex}
    \fi
    \caption[Decoding algorithm's error probabilities]{
        Decoding algorithm's error probabilities.
        The decoding algorithm either returns codeowrds not transmitted ($p_{FA}$) or fails to find transmitted codewords ($p_{MD}$).
        The error probabilities are compared to their theoretical bounds for $N=100$, $\mathcal{M}=1000$, $K=5$, $\rho=24_{\mathrm{dB}}$, $\delta=0.33$, $\gamma^* = 11.318$ and $p^*=0.0986$.
    }
    \label{fig:res_pFA_pMD}
\end{figure}

Figure \ref{fig:res_q10_q01} shows that the calculations in lemmas \ref{eq:CharaErr_q01} and \ref{eq:CharaErr_q10} coincide with the simulation results.
Figure \ref{fig:res_pFA_pMD} shows that propositions \ref{theo:CoMa_pMD_to0_suff_cond} and \ref{theo:CoMa_pFA_to0_suff_cond} hold when taking \mbox{$M_r\geq (1+\delta)\beta^*K\ln N\mathcal{M}$}.
We note that our results are asymptotically tight.

\begin{figure}
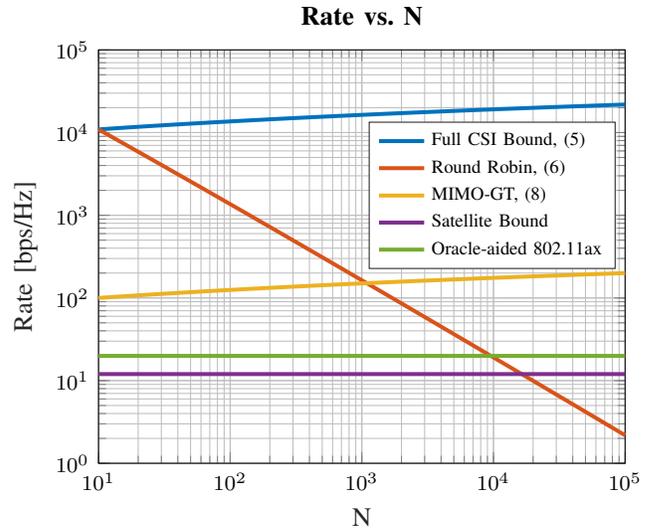

    \centering
    \ifTCOMorTWCDraft
        \input{TexFiles/TexImages/rates_N_TCOM_TWC_draft.tex}
    \else
        \input{TexFiles/TexImages/rates_N}
    \fi
    \caption{
        Different rates (normalized in bandwidth) as functions of $N$ when $\rho=20_{\mathrm{dB}}$, $K=10$, $\mathcal{M}=1000$, $\delta=0.33$ and $N_0 = 2$.
        The 802.11ax is oracle-aided; the $K$ users are scheduled without passing information to the access point.
        The coefficient multiplying the logarithmic term in (\ref{eq:ergodic_Cap_K_CSIT}) is 1.
    }
    \label{fig:rate_ratio_N}
\end{figure}

Figure \ref{fig:rate_ratio_N} compares the rates in Subsection \ref{subsect:Rates} with MIMO-GT's rate.
We also compare our rate with MU systems used in practice - errorless satellite networks (using data provided by AYECKA) and oracle-aided errorless 802.11ax.
The blue line is (\ref{eq:ergodic_Cap_K_CSIT}), whereas the orange line is (\ref{eq:ergodic_Cap_RR}).
The yellow line is the sum-rate of MIMO-GT, (\ref{eq:compare_our_rate}).
The purple line is the sum-rate of raw transmissions at the highest symbol rate of a $K$-to-1 satellite communication (normalized by its bandwidth).

The green line is the sum-rate of a $160$MHz 802.11ax, normalized by a single carrier's bandwidth ($78.125$kHz).
The oracle schedules the $K$ users without passing requests or scheduling information.
I.e., they do not send $\log N$ \emph{header} bits for identification.
The receiver sends a trigger frame followed by a short silent interval (named SIFS).
Afterward, the users transmit their frames simultaneously on different frequency bands.
Payload size is $\log \mathcal{M}$ (for a fair comparison) and is sent in MCS2 rate, and the rest is according to the 802.11ax standard (one training field and $8\mu$sec packet extension).
The sum-rate also considers the trigger frame (sent in MCS0 rate) to allocate resources to the $K$ users; each is scheduled to 200 frequency bands.
The sum-rate can be further reduced when considering the $\log N$ header bits and the scheduling information.
MIMO-GT achieves higher sum-rates than 802.11ax (when the number of users is big enough) or satellite networks and has no significant overheads.

\begin{figure}
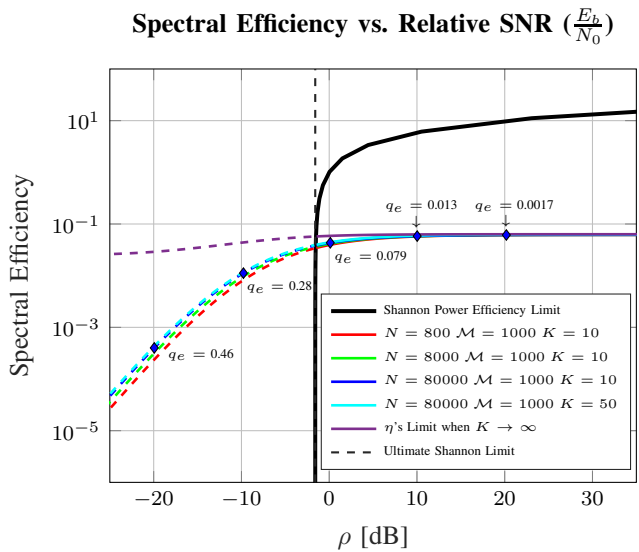

    \centering
    \ifTCOMorTWCDraft
        \input{TexFiles/TexImages/spec_eff_TCOM_TWC_draft.tex}
    \else
        \input{TexFiles/TexImages/spec_eff}
    \fi
    \caption{
        Spectral Efficiency as a function of $\rho$ (in $\mathrm{dB}$) when $\delta=0.33$ and $N_0 = 2$.
        The mentioned probabilities, $q_e\triangleq(1-p)^K q_{01} + (1-(1-p)^K) q_{10}$, are the BER of the blue line at the corresponding SNR.
    }
    \label{fig:spec_eff}
\end{figure}

Figure \ref{fig:spec_eff} compares MIMO-GT's Spectral Efficiency from (\ref{eq:compare_our_spec_eff}) with SPEL, evaluated for different settings of $K$, $N$ and $\mathcal{M}$.
The SPEL is calculated like in \cite[Chapter~3.5]{srinivasan2020wireless}.
For our system, we have used $\frac{E_b}{N_0} = \frac{K P}{N_0 M_r}$.
The bold dark line is SPEL, and its dashed counterpart is the \emph{absolute} SPEL, $\ln 2$.
The purple line is $\eta$'s (eq. (\ref{eq:compare_our_spec_eff})) limit when $K\to\infty$, calculated regardless of $N$ or $\mathcal{M}$.

\begin{figure}
    \centering
    \input{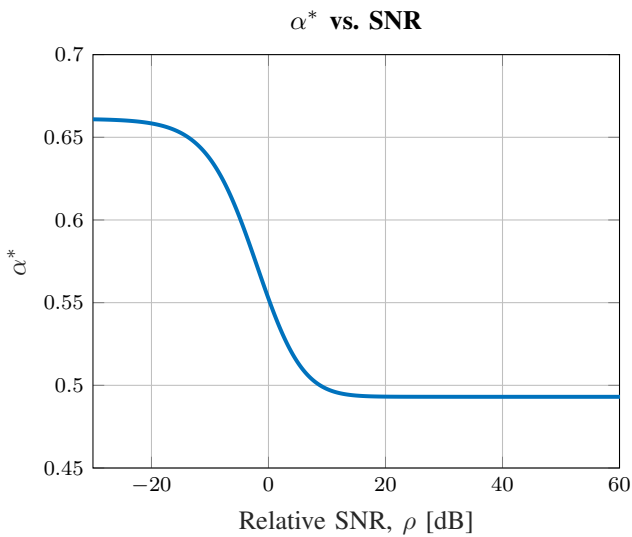}
    \caption{
        $\alpha^* \triangleq K p^*$'s value as a function of $\rho$ when $N=100$, $K=5$, $\mathcal{M}=1000$, $\delta=0.33$ and $N_0 = 2$.
    }
    \label{fig:alpha_f_rho}
\end{figure}

\begin{figure}
    \centering
    \input{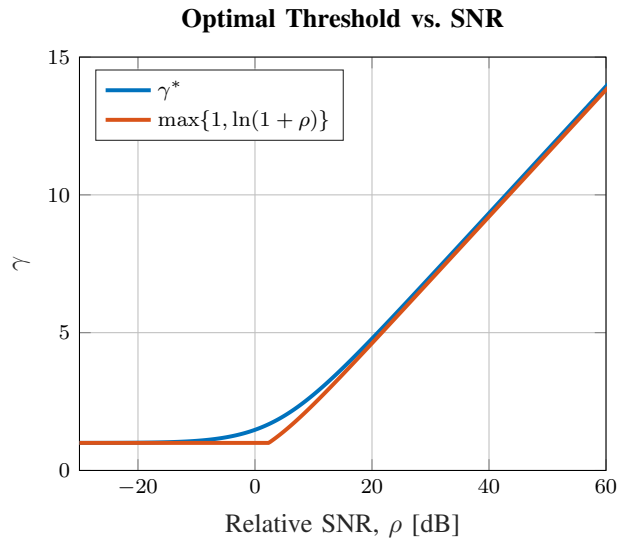}
    \caption{
        $\gamma^*$'s value as a function of $\rho$ when $N=100$, $K=5$, $\mathcal{M}=1000$, $\delta=0.33$ and $N_0 = 2$ compared to the heuristic $\gamma = \max\{1, \ln (1+\rho)\}$.
    }
    \label{fig:gamma_f_rho}
\end{figure}

\begin{figure}
    \centering
    \input{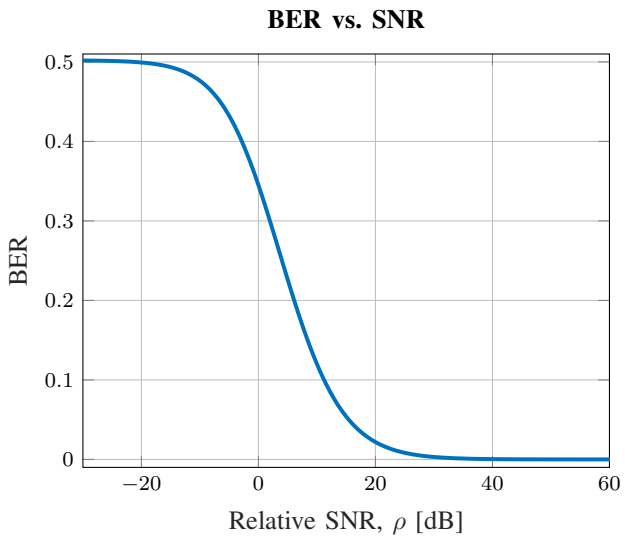}
    \caption{
        BER as a function of $\rho$ when $N=100$, $K=5$, $\mathcal{M}=1000$, $\delta=0.33$ and $N_0 = 2$.
    }
    \label{fig:BER_f_rho}
\end{figure}

Figures \ref{fig:alpha_f_rho}, \ref{fig:gamma_f_rho} and \ref{fig:BER_f_rho} show how $\alpha^*\triangleq K p^*$, $\gamma^*$, and the BER\footnote{ evaluated as $q_e\triangleq(1-p)^K q_{01} + (1-(1-p)^K) q_{10}$} (respectively) scale as a function of $\rho\in[-30, 60]_{\mathrm{dB}}$.
When the SNR is low, the BER tends to 0.5, equivalent to the error when the receiver guesses whether the antenna is activated.
$\gamma^*$ tends to 1 when the SNR is low, unlike its proportion to $\ln \rho$ in the high SNR region.
Since $\gamma$ can be independent of the code we use, choosing $\gamma = \max\{1, \ln (1+\rho)\}$ is an excellent \emph{heuristic}.

\section{Conclusion}
\label{sect:conc}
In this paper, we studied a distributed MU-MIMO scheme using GT codes on the antennas at the receiver where the users are non-cooperative self-scheduling named MIMO-GT.
The receiver used energy detection in each antenna and a simple decoding algorithm to jointly obtain numerous messages.
Our approach is simple to implement and order-optimal in the number of users or messages.
MIMO-GT's order-optimality is shown by comparing our rate to the Full CSI solution, and the number of antennas required asymptotically matches the converse bound we calculated.
We have expressed and determined the scaling laws of the antennas when the SNR or the number of users grows large.
Our results are supported by simulations and numerical evaluations (e.g., matching slopes between MIMO-GT's rate and the Full CSI bound).

MIMO-GT relies heavily on the reliability of the channel estimation at \emph{each} transmitter; if a user errs in their estimation, the communication may fail.
Future research may address this issue by studying the effects of estimation errors or aiming for different algorithms to jointly obtain the sent messages without utilizing the perfectly estimated channel matrices.
We may overcome the errors by either using a special deterministic codebook (with a constant number of intersections between every $K$ codewords), by devising a scheme non-reliant on CSIT (e.g., utilizing CSIR), or by adding more antennas to compensate for the erroneous channel estimation.
The codebooks of the first approach are hard to find, and the last approach must be designed meticulously to not accidentally change the antenna scaling laws, potentially annihilating the order-optimality of MIMO-GT.
The second approach might be more practical as excellent CSI collection techniques are available at the receiver.
On the other hand, not utilizing CSI at all results in massive rate loss.

% \section*{Acknowledgment}
% \label{sect:ACK}
% \pageref{sect:ACK}
% The authors thank AYECKA for providing information about possible satellite rates.

\bibliographystyle{IEEEtran}
\bibliography{references}
%% Appendix
% \newpage
\appendices
\section{Proofs}
\label{appendix:proofs}
\subsection{Proof of Lemma \ref{theo:CoMa_b_opt_prob_equiv}}
\label{appendix:proof_CoMa_b_opt_prob_equiv}
First, we simplify $\frac{p_0}{q_{10}}-1$ to
\ifTCOMorTWCDraft
    \begin{align}
        \label{eq:CoMa_Delta_up_simp}
        \frac{p_0}{q_{10}} -1
        &=
        \frac{(1-p)^K (1-q_{01}) + (1-(1-p)^K)q_{10}}{q_{10}} - 1
        =
        (1-p)^K \frac{1-q_{01}-q_{10}}{q_{10}}.
    \end{align}
\else
    \begin{align}
        \nonumber
        \frac{p_0}{q_{10}} -1
        &=
        \frac{(1-p)^K (1-q_{01}) + (1-(1-p)^K)q_{10}}{q_{10}} - 1
        \\\label{eq:CoMa_Delta_up_simp}
        &=
        (1-p)^K \frac{1-q_{01}-q_{10}}{q_{10}}.
    \end{align}
\fi

Now, we convert the problem in (\ref{eq:CoMa_b_opt_prob_orig}) to a minimization problem by removing the dependence on $\Delta$.
\begin{proposition}
    \label{eq:CoMa_b_opt_prob_orig_reduce_minimax}
    The objective function in (\ref{eq:CoMa_b_opt_prob_orig}) can be re-written as follows:
    \begin{align}
        \nonumber
        \frac{1}{K p\big(1-\exp\{-\frac{1}{2}(1-p)^{2K}(1-q_{10}-q_{01})^2\}\big)}
    \end{align}
\end{proposition}
\begin{proof}
    By deriving $\beta_i$ by $\Delta$, we obtain
    \begin{align}
        \frac{\partial\beta_1}{\partial \Delta}
        &=
        -\frac{4 q_{10}^2 e^{-2(q_{10}\Delta)^2}\Delta}{(\cdot)^2} \leq 0 \ \forall \Delta\geq0
        \\
        \frac{\partial\beta_2}{\partial \Delta}
        &=
        \frac{4q_{10}e^{-2(p_0-q_{10}(\Delta+1))^2}(p_0-q_{10}(\Delta+1))}{(\cdot)^2}.
    \end{align}
    The last term is non-negative as long as $\Delta\leq \frac{p_0}{q_{10}}-1$.
    Hence, $\beta_1$ decrease with $\Delta$ whereas $\beta_2$ increase in $\Delta$.
    In other words, the minimax problem would pick the equalizer, 
    \begin{align}
        \Delta^*
        &=
        \frac{1}{2}\bigg(\frac{p_0}{q_{10}}-1\bigg).
    \end{align}
    Putting $\Delta^*$ and (\ref{eq:CoMa_Delta_up_simp}) back into $\beta_1=\beta_2$ results in the following term inside the exponent
    \begin{align}
        q_{10} \frac{1}{2}\bigg(\frac{p_0}{q_{10}}-1\bigg)
        &=
        \frac{1}{2}(1-p)^K(1-q_{01}-q_{10}).
    \end{align}
\end{proof}

Now, we define $p \triangleq \frac{\alpha}{K}$.
Assuming $\alpha^*$ exists, we shall show that $\alpha^*\in (0,\frac{K}{2})$.
That is, $\alpha^*$ is an interior point.
\begin{proposition}
    \label{lem:eq:CoMa_b_opt_prob_orig_alphaStar_bounded}
    If $\alpha^*$ exists, $\alpha^*\leq e-1$
\end{proposition}
\begin{proof}
    Let us observe the objective function in proposition \ref{eq:CoMa_b_opt_prob_orig_reduce_minimax} derivative's numerator as a function of $\alpha$, denoted as $g(\alpha)$.
    \ifTCOMorTWCDraft
        \begin{align}
            \nonumber
            g(\alpha)
            &=
            \bigg\{
                -1+e^{-\frac{1}{2}(1-p)^{2K}(1-q_{10}-q_{01})^2}
                +
                \alpha e^{-\frac{1}{2}(1-p)^{2K}(1-q_{10}-q_{01})^2}\bigg(1-\frac{\alpha}{K}\bigg)^{2K-1}
                \\\nonumber
                &\qquad\cdot (1-q_{01}-q_{10})\bigg(
                    1-q_{01}-q_{10}
                    +
                    \bigg(
                        1-\frac{\alpha}{K}
                    \bigg)
                    \frac{\partial q_{10}}{\partial \alpha}
                \bigg)
            \bigg\}.
        \end{align}
    \else
        \begin{align}
            \nonumber
            g(\alpha)
            &=
            \bigg\{
                -1+e^{-\frac{1}{2}(1-p)^{2K}(1-q_{10}-q_{01})^2}
                +
                \\\nonumber
                &\qquad
                +\alpha e^{-\frac{1}{2}(1-p)^{2K}(1-q_{10}-q_{01})^2}\bigg(1-\frac{\alpha}{K}\bigg)^{2K-1}
                \\\nonumber
                &\qquad\cdot (1-q_{01}-q_{10})\bigg(
                    1-q_{01}-q_{10}
                    +
                    \\
                    &\qquad
                    +
                    \bigg(
                        1-\frac{\alpha}{K}
                    \bigg)
                    \frac{\partial q_{10}}{\partial \alpha}
                \bigg)
            \bigg\}.
        \end{align}
    \fi
    In other words, $g(\alpha)$ is of the form
    \begin{align}
        g(\alpha)
        =
        c_1 (e^{-h(\alpha)}(1+\alpha w(\alpha)) - 1).
    \end{align}
    Where $w(\alpha) = (1-\alpha/K)^{2K-1} (1-q_{01}-q_{10}) \big(1-q_{01}-q_{10}+(1-\alpha/K)\frac{\partial q_{10}}{\partial \alpha}\big)$, $h(\alpha) = \frac{1}{2}(1-p)^{2K}(1-q_{10}-q_{01})^2$ and $c_1 > 0$.
    $g(\alpha^*) = 0$ if and only if $1+\alpha^* w(\alpha^*) = e^{h(\alpha^*)}$.
    We notice that $0\leq h(\alpha^*)\leq 1$, so
    \begin{align}
        e^{0} \leq 1+\alpha^* w(\alpha^*) \leq e^{1}
        \\
        0 \leq \alpha^* w(\alpha^*) \leq e-1
    \end{align}
    Observe $\frac{\partial q_{10}}{\partial p}$
    \ifTCOMorTWCDraft
        \begin{align}
            \nonumber
            \frac{\partial q_{10}}{\partial p}
            &=
            \sum_{j=1}^K \frac{\binom{K}{j}p^j(1-p)^{K-j}}{(1-(1-p)^K)^2 p (1-p)}(1-e^{-\frac{\gamma}{1+j\rho}})  (j(1-(1-p)^K)-K p).
        \end{align}
    \else
        \begin{align}
            \nonumber
            \frac{\partial q_{10}}{\partial p}
            &=
            \sum_{j=1}^K \frac{\binom{K}{j}p^j(1-p)^{K-j}}{(1-(1-p)^K)^2 p (1-p)}\cdot
            \\
            &\qquad
            \cdot(1-e^{-\frac{\gamma}{1+j\rho}})  (j(1-(1-p)^K)-K p).
        \end{align}
    \fi
    Since $K \geq 2$, $\alpha > (1-(1-\alpha/K)^K)$ any $\alpha\geq 0$, so $\frac{\partial q_{10}}{\partial \alpha} < 0$.
    Therefore $w(\alpha) \leq 1$ and we conclude that $\alpha^*\leq e-1$.
\end{proof}

Now we will concentrate on proving that $\gamma^*$ exists.
We notice that minimizing the objective function in proposition \ref{eq:CoMa_b_opt_prob_orig_reduce_minimax} as a function of $\gamma$ is the same as minimizing $f(\gamma)\triangleq q_{01}+q_{10}$.
As a result, proving $\gamma^*$'s existence requires showing it exists for $f(\gamma)$.
\begin{proposition}
    \label{eq:CoMa_b_opt_prob_orig_gammaStar_exists}
    For any $\alpha < K$, there exists a \emph{unique} $\gamma^*$ to the optimization problem.
\end{proposition}
We want to remark that despite what proposition \ref{eq:CoMa_b_opt_prob_orig_gammaStar_exists} might suggest, $f(\gamma)$ is not convex nor concave in $\gamma\in[0,\infty)$.
\emph{Proof Sketch}:
We fist show that there exists some $\gamma_0$ such that $\frac{\partial f}{\partial\gamma} = 0$.
Next, by explicitly writing $\frac{\partial f}{\partial\gamma}$, we notice that the terms inside the exponents are linear in $\gamma$, so $\gamma_0$ must be unique.
Next, we calculate $\frac{\partial^2 f}{\partial^2\gamma}\big|_{\gamma = \gamma_0}$ to learn that it's strictly positive.
I.e., $\gamma_0$ minimizes $f(\gamma)$ so $\gamma_0 = \gamma^*$.
\begin{proof}
    $q_{10}$ is well defined when $\alpha < K$.
    $f(0) = 1$ from lemmas \ref{eq:CharaErr_q01} and \ref{eq:CharaErr_q10}.
    When $\gamma\to\infty$, $f(\gamma)\to 1$, so by Rolle's Theorem\footnote{formally, there exists $y_0 < 1$, $\gamma_1$ and $\gamma_2$ such that $f(\gamma_1)=f(\gamma_2)=y_0$.} there exists some $\gamma_0$ such that $\frac{\partial f}{\partial\gamma}\big|_{\gamma=\gamma_0} = 0$.
    
    Let us calculate $\frac{\partial f}{\partial\gamma}$ explicitly:
    \begin{align}
        \label{eq:gammaStar_helper1}
        \frac{\partial f}{\partial\gamma}
        =
        -e^{-\gamma} + \sum_{j=1}^K \binom{K}{j}\frac{p^j(1-p)^{K-j}\exp\big\{-\frac{\gamma}{j \rho + 1}\big\}}{(1-(1-p)^K)(1+j\rho)}.
    \end{align}
    All the terms inside the exponents are linear in $\gamma$, so $\gamma_0$ is unique.
    Now we are ready to use the second derivative test.
    \begin{align}
        \label{eq:gammaStar_helper2}
        \frac{\partial^2 f}{\partial^2\gamma}
        =
        e^{-\gamma} - \sum_{j=1}^K \binom{K}{j}\frac{p^j(1-p)^{K-j}\exp\big\{-\frac{\gamma}{j \rho + 1}\big\}}{(1-(1-p)^K)(1+j\rho)^2}.
    \end{align}
    $\gamma_0$ nulls (\ref{eq:gammaStar_helper1}), so we can replace $e^{-\gamma}$ in (\ref{eq:gammaStar_helper2}) with the sum in (\ref{eq:gammaStar_helper1}) to obtain a sum of differences
    \begin{align}
        \sum_{j=1}^K \binom{K}{j}\frac{p^j(1-p)^{K-j}}{1-(1-p)^K} e^{-\frac{\gamma_0}{j \rho + 1}} \bigg(\frac{1}{1+j\rho}-\frac{1}{(1+j\rho)^2}\bigg).
    \end{align}
    Hence, the sign of the second derivative is decided by the sign of differences
    \begin{align}
        \frac{1}{1+j\rho} - \frac{1}{(1+j\rho)^2}
        =
        \frac{1+j\rho-1}{(1+j\rho)^2}
        =
        \frac{j\rho}{(1+j\rho)^2} > 0.
    \end{align}
    In other words, each addend is positive, so $\gamma_0$ minimizes $f(\gamma)$.
\end{proof}

Now we are ready to prove lemma \ref{theo:CoMa_b_opt_prob_equiv}.
    \begin{proof}[Proof of Lemma \ref{theo:CoMa_b_opt_prob_equiv}]
    Since $[0, \frac{K}{2}]$ is a closed interval, and the objective function is continuous, $\alpha^*$ (which brings the objective function in proposition \ref{eq:CoMa_b_opt_prob_orig_reduce_minimax} to a global minimum) exists.
    Proposition \ref{lem:eq:CoMa_b_opt_prob_orig_alphaStar_bounded} bounds $\alpha^*$ to the interval $(0, e-1]$.
    Proposition \ref{eq:CoMa_b_opt_prob_orig_gammaStar_exists} assures that $\gamma^*$ exists and is unique per $\alpha$ (and $\alpha^*$ in particular), so a unique solution to the optimization problem exists (consequently, $p^*$ exists).
    We also notice that $\gamma^*$ is the solution to the optimization problem in equation (\ref{eq:CoMa_b_opt_prob_orig}), so $(\gamma^*, p^*)$ are the solutions to optimization problem in equation (\ref{eq:CoMa_b_opt_prob_orig}).
\end{proof}

To ease the numerical search of $\gamma^*$ and $p^*$ numerically, we show that closing the intervals does not introduce new solutions to the optimization problem:
\begin{proposition}
    \label{eq:CoMa_b_opt_prob_orig_reduce_weak_ineq}
    Rewriting all strong inequalities in (\ref{eq:CoMa_b_opt_prob_orig}) with weak inequalities does not introduce new solutions.
    Additionally, extending the search of $p^*$ to the interval $[0,0.5]$ does not introduce new solutions.
\end{proposition}
\begin{proof}
    $q_{10} = 0$ when $\gamma = 0$ so $\beta_1\to\infty$.
    When $p = 0$ both $\beta_1$ and $\beta_2$ tend to infinity.
    Hence, we can replace the strong inequalities with weak inequalities.
    Proposition \ref{lem:eq:CoMa_b_opt_prob_orig_alphaStar_bounded} assures $p^* = O(\frac{1}{K})$ and proposition \ref{eq:CoMa_b_opt_prob_orig_gammaStar_exists} assures it is unique, so extending $p$'s range does not include new solutions.
\end{proof}

\subsection{Proof of Proposition \ref{lem:betaStar:bound}}
\label{appendix:proof:betaStar:bound}
$\beta^*$ is the solution of lemma \ref{theo:CoMa_b_opt_prob_equiv}'s optimization problem, so it is enough to show that 
\begin{align}
    \nonumber
    1-e^{-\frac{1}{2}(1-p)^{2K}(1-q_{10}-q_{01})^2}
    \geq
    \frac{3(1-\frac{\alpha}{2})^4 \rho^2}{8 e^{2\max\{\rho,1\}} (\rho+1)^2}.
\end{align}
We first bound $q_{10}$ by its largest addend.
That is, 
\begin{align}
    q_{10} \leq 1-e^{\frac{-\gamma}{\rho + 1}}.
\end{align}
Hence, the terms inside the exponent are bounded by
\ifTCOMorTWCDraft
    \begin{align}
        1-q_{01}-q_{10}
        &\geq
        e^{\frac{-\gamma}{\rho + 1}}
        -
        e^{-\gamma}
        =
        e^{-\gamma} \big(
            e^{\frac{\gamma\rho}{\rho +1}}-1
        \big)
        \\\label{eq:betaStar:proof:A}
        &=    
        e^{-\gamma} \bigg(
            1+\frac{\gamma\rho}{\rho +1}+O\bigg(\bigg(\frac{\gamma\rho}{\rho +1}\bigg)^2\bigg)-1
        \bigg)
        \\\label{eq:betaStar:proof:B}
        &\geq
        e^{-\gamma}\frac{\gamma\rho}{\rho +1}
        \geq
        \frac{\rho}{\rho +1}e^{-\max\{1,\rho\}}
    \end{align}
\else
    \begin{align}
        1-q_{01}-q_{10}
        &\geq
        e^{\frac{-\gamma}{\rho + 1}}
        -
        e^{-\gamma}
        \\
        &=
        e^{-\gamma} \big(
            e^{\frac{\gamma\rho}{\rho +1}}-1
        \big)
        \\\label{eq:betaStar:proof:A}
        &=    
        e^{-\gamma} \bigg(
            1+\frac{\gamma\rho}{\rho +1}+O\bigg(\bigg(\frac{\gamma\rho}{\rho +1}\bigg)^2\bigg)-1
        \bigg)
        \\
        &\geq
        e^{-\gamma}\frac{\gamma\rho}{\rho +1}
        \\\label{eq:betaStar:proof:B}
        &\geq
        \frac{\rho}{\rho +1}e^{-\max\{1,\rho\}}
    \end{align}
\fi
In equation (\ref{eq:betaStar:proof:A}) we used the Taylor series of $e^{-x} = 1-x+O(x^2)$. %$\frac{1}{2}x^2-O(x^3)$.
The last step used $\gamma$'s range in the proposition.
Now, we bound $(1-p)^{2K}$ with the famous limit 
\begin{align}
    \label{eq:beta_scale_limit1}
    \bigg(
        1-\frac{\alpha}{K}
    \bigg)^{K}
    \to
    e^{-\alpha}.
\end{align}
The bounds are
\begin{align}
    \label{eq:betaStar:proof:C}
    \bigg(1-\frac{\alpha}{2}\bigg)^4
    \leq
    \bigg(
        1-\frac{\alpha}{K}
    \bigg)^{2K}
    \leq
    e^{-2\alpha}.
\end{align}
By putting (\ref{eq:betaStar:proof:B}), (\ref{eq:betaStar:proof:C}) and taking additional addend in the Taylor series of $e^{-x} = 1-x+\frac{1}{2}x^2-O(x^3)$ we obtain that
\ifTCOMorTWCDraft
    \begin{align}
        1-&e^{-\frac{1}{2}(1-p)^{2K}(1-q_{10}-q_{01})^2}
        \geq
        1-\exp\bigg\{-\frac{1}{2}\bigg(1-\frac{\alpha}{2}\bigg)^{4}\bigg(\frac{\rho}{\rho +1}\bigg)^2e^{-2\max\{1,\rho\}}\bigg\}
        \\
        &\geq
        \frac{1}{2}\bigg(1-\frac{\alpha}{2}\bigg)^{4}\bigg(\frac{\rho}{\rho +1}\bigg)^2 e^{-2\max\{1,\rho\}}
        -
        \frac{1}{8}\bigg(1-\frac{\alpha}{2}\bigg)^{8}\bigg(\frac{\rho}{\rho +1}\bigg)^4 e^{-4\max\{1,\rho\}}
        \\
        &\geq
        \frac{1}{2}\bigg(1-\frac{\alpha}{2}\bigg)^{4}\bigg(\frac{\rho}{\rho +1}\bigg)^2 e^{-2\max\{1,\rho\}}
        -
        \frac{1}{8}\bigg(1-\frac{\alpha}{2}\bigg)^{4}\bigg(\frac{\rho}{\rho +1}\bigg)^2 e^{-2\max\{1,\rho\}}
        \\
        &\geq
        \frac{3}{8}\bigg(1-\frac{\alpha}{2}\bigg)^{4}\bigg(\frac{\rho}{\rho +1}\bigg)^2 e^{-2\max\{1,\rho\}}
    \end{align}
\else
    \begin{align}
        \nonumber
        &1-e^{-\frac{1}{2}(1-p)^{2K}(1-q_{10}-q_{01})^2}
        \\
        &
        \geq
        1-\exp\bigg\{-\frac{1}{2}\bigg(1-\frac{\alpha}{2}\bigg)^{4}\bigg(\frac{\rho}{\rho +1}\bigg)^2e^{-2\max\{1,\rho\}}\bigg\}
        \\\nonumber
        &\geq
        \frac{1}{2}\bigg(1-\frac{\alpha}{2}\bigg)^{4}\bigg(\frac{\rho}{\rho +1}\bigg)^2 e^{-2\max\{1,\rho\}}
        \\
        &\qquad
        -
        \frac{1}{8}\bigg(1-\frac{\alpha}{2}\bigg)^{8}\bigg(\frac{\rho}{\rho +1}\bigg)^4 e^{-4\max\{1,\rho\}}
        \\\nonumber
        &\geq
        \frac{1}{2}\bigg(1-\frac{\alpha}{2}\bigg)^{4}\bigg(\frac{\rho}{\rho +1}\bigg)^2 e^{-2\max\{1,\rho\}}
        \\
        &\qquad
        -
        \frac{1}{8}\bigg(1-\frac{\alpha}{2}\bigg)^{4}\bigg(\frac{\rho}{\rho +1}\bigg)^2 e^{-2\max\{1,\rho\}}
        \\\nonumber
        &\geq
        \frac{3}{8}\bigg(1-\frac{\alpha}{2}\bigg)^{4}\bigg(\frac{\rho}{\rho +1}\bigg)^2 e^{-2\max\{1,\rho\}}
    \end{align}
\fi

\subsection{Proof of Lemma \ref{lem:beta_scale_K_inf}}
\label{appendix:proof_beta_scale_K_inf}
Since $K p = \alpha$, $\beta^*$'s convergence to a constant depends on the limit
\ifTCOMorTWCDraft
    \begin{align}
        \lim_{K\to\infty} q_{10}
        &=
        \lim_{K\to\infty}\sum_{j=1}^K \binom{K}{j}\frac{p^j(1-p)^{K-j}}{1-(1-p)^K}\bigg(1 - \exp\bigg\{\frac{-\gamma}{j \rho + 1}\bigg\}\bigg)
        \\\label{eq:q10Limit:proof:A}
        &=
        1-\lim_{K\to\infty}\sum_{j=1}^K \binom{K}{j}\frac{p^j(1-p)^{K-j}}{1-(1-p)^K}\exp\bigg\{\frac{-\gamma}{j \rho + 1}\bigg\}
        \\\label{eq:q10Limit:proof:B}
        &=
        1-\lim_{K\to\infty}\frac{1}{1-(1-p)^K}\lim_{K\to\infty}\sum_{j=1}^K \binom{K}{j}p^j(1-p)^{K-j}\exp\bigg\{\frac{-\gamma}{j \rho + 1}\bigg\}
        \\\label{eq:beta_scale_limit2}
        &=
        1-\frac{1}{1-e^{-\alpha}} \sum_{j=1}^\infty e^{-\alpha}\frac{\alpha^j}{j!}\exp\bigg\{\frac{-\gamma}{j \rho + 1}\bigg\}
    \end{align}
\else
    \begin{align}
        \nonumber
        \lim_{K\to\infty} &q_{10}
        =
        \\
        &\lim_{K\to\infty}\sum_{j=1}^K \binom{K}{j}\frac{p^j(1-p)^{K-j}}{1-(1-p)^K}
        % \\\nonumber
        % &\qquad\cdot
        \bigg(1 - \exp\bigg\{\frac{-\gamma}{j \rho + 1}\bigg\}\bigg)
    \end{align}
    \begin{align}
        \label{eq:q10Limit:proof:A}
        &=
        1-\lim_{K\to\infty}\sum_{j=1}^K \binom{K}{j}\frac{p^j(1-p)^{K-j}}{1-(1-p)^K}
        \exp\bigg\{\frac{-\gamma}{j \rho + 1}\bigg\}
        \\\nonumber
        &=
        1-\lim_{K\to\infty}\frac{1}{1-(1-p)^K}
        \\\label{eq:q10Limit:proof:B}
        &\quad
        \cdot\lim_{K\to\infty}\sum_{j=1}^K \binom{K}{j}p^j(1-p)^{K-j}\exp\bigg\{\frac{-\gamma}{j \rho + 1}\bigg\}
        \\\label{eq:beta_scale_limit2}
        &=
        1-\frac{1}{1-e^{-\alpha}} \sum_{j=1}^\infty e^{-\alpha}\frac{\alpha^j}{j!}\exp\bigg\{\frac{-\gamma}{j \rho + 1}\bigg\}
    \end{align}
\fi
In equations (\ref{eq:q10Limit:proof:A}) and (\ref{eq:q10Limit:proof:B}) we used the Limit Sum and Product Laws, respectively.
The final step, (\ref{eq:beta_scale_limit2}), used Poisson Limit Theorem and the famous limit.
The series in equation (\ref{eq:beta_scale_limit2}) is convergent by d'Alembert's criterion (with respect to $e^{x}$'s Taylor series).
\\
Combining (\ref{eq:beta_scale_limit1}) and (\ref{eq:beta_scale_limit2}) with lemma \ref{theo:CoMa_b_opt_prob_equiv} when $K\to\infty$ results in an optimization problem independent of $K$, $N$, or $\mathcal{M}$, completing the proof.

\end{document}